\documentclass{emulateapj}
\def\farcs{\hbox{$~\mkern-4mu^{\prime\prime}$}}

\def\la{\mathrel{\hbox{\rlap{\hbox{\lower4pt\hbox{$\sim$}}}\hbox{$<$}}}}
\def\ga{\mathrel{\hbox{\rlap{\hbox{\lower4pt\hbox{$\sim$}}}\hbox{$>$}}}}

\shortauthors{Alan}
\shorttitle{SNR 1E 0102.2-7219}

\begin{document}

\title{A Detailed Archival $\it{CHANDRA}$ Study of the Young Core-Collapse \\ Supernova Remnant 1E 0102.2-7219 in the Small Magellanic Cloud}

\author{Neslihan Alan\altaffilmark{1,2,3}, Sangwook Park\altaffilmark{2}, Sel\c{c}uk Bilir\altaffilmark{1}}
\altaffiltext{1}{Istanbul University, Faculty of Science, Department of Astronomy and Space Sciences, 34119, Beyazit, Istanbul, Turkey; neslihan.alan@gmail.com; sbilir@istanbul.edu.tr }
\altaffiltext{2}{Department of Physics, University of Texas at Arlington, Arlington, TX 76019; s.park@uta.edu}
\altaffiltext{3}{Visiting scholar at UTA}

\begin{abstract}

We present an archival {\it Chandra} study of the O-rich supernova remnant (SNR) 1E 0102.2-7219 in the Small Magellanic Cloud. Based on the deep $\sim$ 265 ks archival {\it Chandra} data we performed a detailed spatially resolved spectral analysis of 1E 0102.2-7219. Our aim is to reveal the spatial and chemical structures of this remnant in unprecedented details. Radial profiles of O, Ne and Mg abundances based on our analysis of regional spectra extracted along nine different azimuthal directions of 1E 0102.2-7219 suggest the contact discontinuity at $\sim$ 5--5.5 pc from the geometric center of the X-ray emission of the SNR. We find that the metal-rich ejecta gas extends further outward in west and southwest than in other directions of the SNR. The average O/Ne, O/Mg and Ne/Mg abundance ratios of the ejecta are in plausible agreement with the nucleosynthesis products from the explosion of a $\sim 40 M_\odot$ progenitor. We estimate an upper limit on the Sedov age of $\sim 3500$ yr and explosion energy of $\sim 1.8\times10^{51}$ erg for 1E 0102.2-7219. We discuss the implications of our results on the geometrical structure of the remnant, its circumstellar medium and the nature of the progenitor star. Our results do not fit with a simple picture of the reverse-shocked emission from a spherical shell-like ejecta gas with a uniformly-distributed metal abundance and a power-law density along the radius of the SNR. 

\end{abstract}

\section {\label{sec:intro} INTRODUCTION}

Supernova remnants (SNRs) are vital to reveal how stars synthesize and disseminate elements heavier than H and He. Core-collapse explosions of massive stars ($M > 8 M_{\odot}$) account for $\sim 3/4$ of all supernovae \citep[][]{tsujimoto95, sato07}. Young core-collapse supernova remnants with uncontaminated debris are among our best natural laboratories for understanding how nucleosynthesis operates in massive stars. Studying X-ray data of core-collapse supernova remnants  provides clues about the dynamics of supernova explosion, the chemical composition of progenitor and the circumstellar medium (CSM), and ultimately about the chemical enrichment process in galaxies. 

1E 0102.2-7219 (E0102), the youngest and brightest supernova remnant in the Small Magellanic Cloud (SMC), was discovered by  {\it Einstein} satellite  \citep{seward81}. \cite{vogt10} mapped the [O III] $\lambda$5007 dynamics of the young oxygen-rich supernova remnant using data of the Wide Field Spectrograph on the 2.3 m telescope at Siding Spring Observatory and they deduced that the ejecta trace an asymmetric bipolar structure. There are different estimations for the age of E0102. \cite{tuohy83} kinematically estimated $\sim 1000$ years for the age of the SNR using high-velocity oxygen-rich optical filaments in the remnant. \cite{hughes00} calculated the expansion rate of the remnant using {\it Einstein}, {\it ROSAT}, and {\it Chandra} X-ray images. The expansion rate calculated by \cite{hughes00} inferred an age of $\sim 1000$ years for E0102. \cite{finkelstein06} estimated a mean expansion velocity for bright ejecta of $\sim 2000$ km s$^{-1}$ from proper motion measurements of 12 optical ejecta filaments and they calculated an age of $\sim 2000$ years for E0102. The progenitor mass of E0102 is uncertain. \cite{flanagan04} calculated the mass of oxygen in the ejecta as 6 $M_\odot$\ using {\it Chandra} X-ray data and they determined total mass of the progenitor as 32 $M_\odot$. \cite{finkelstein06} suggested that the progenitor mass of E0102 is larger than 50 $M_\odot$\, taking into account the masses of Wolf-Rayet and O-type stars in the HII region N76A located in the southwest of E0102 \citep{massey00}. It was known that E0102 is originated from the core-collapse because of its ``oxygen-rich'' nature observed in the optical band \citep{dopita81}. \cite{rutkowski10} did not detect the central point source near the X-ray geometric center of E0102. Recently, \cite{vogt18} reported the detection of a candidate central compact object in E0102 using integral field spectroscopy observations from the Multi Unit Spectroscopic Explorer at the Very Large Telescope and {\it Chandra} X-ray Observatory data.

The brightest emission feature in 0.3--7 keV energy range is the ring-like (in projection) metal-rich ejecta in which the X-ray emission is dominated by enhanced lines from highly ionized O, Ne, and Mg \citep{hayashi94, gaetz00}. X-ray emission from the shocked ambient medium is faint, extending beyond the bright ejecta ring \citep{hughes00}. Since the inner boundary of the ring shows a sharp decline in its intensity, \citet{hughes94} suggested that a true ring-like structure rather than a projected spherical shell. Based on {\it Chandra} HETG observations, \cite{flanagan04} found that a simple 3-D model of a nonuniform spherical shell with emission concentrated toward the equatorial plane can explain some, but not all, of the observed features. The ejecta ring might have been created by a non-spherically symmetric supernova explosion \citep{flanagan04} or by an expanding Fe-Ni bubble \citep{eriksen01}. While the presence of metal-rich ejecta and its progressive ionization by the reverse shock in the bright ring were revealed based on qualitative assessments of flux strengths and radial profiles of individual lines \citep{gaetz00}, a quantitative, detailed spatially-resolved spectral analysis of the ejecta has not been performed up to date. Even though less-spatially-resolved X-ray studies have been accomplished previously \citep[][]{sasaki06, sasaki01, rasmussen01, hayashi94}, detailed spatial features of the remnant have not been determined. Moreover, the spectral study of the faint forward shock emission extending beyond the ejecta ring is essential to reveal the dynamics of the SNR. The thermal forward-shocked emission is also essential to estimate the nature of the CSM (e.g., abundances), but such a study was not feasible in previous works because of limited photon statistics due to either a short exposure in the ACIS data \citep[40 ks,][]{gaetz00} or the low sensitivity in the HETG data \citep{flanagan04}. An extensive spatially-resolved X-ray spectral analysis of E0102 is essential to reveal its spatial and chemical structures, and thus its detailed nature of the progenitor, explosion, and SNR dynamics. E0102 has been a calibration source for {\it Chandra} since 1999, for this reason, the {\it Chandra} archive presents a wealth of data for the remnant \citep[][]{plucinsky17}. The {\it Chandra} ACIS provides the imaging spectroscopy with the superb spatial resolution ($\sim0.5\farcs$), which is required for the extensive spatially-resolved spectral analysis of E0102.

\section{\label{sec:obs} OBSERVATIONS \& DATA REDUCTION}

We used deep ($\sim$273\ ks) {\it Chandra} archival data of E0102, a total of 20 ObsIDs which were obtained between 1999 August 23 and 2012 January 12. We summarize these {\it Chandra} observations in Table 1. The X-ray spectrum of E0102 is dominated by emission in the soft X-ray band \citep[$E \la 1$ keV,][]{gaetz00}. The overall angular size of E0102 is small ($\sim 44\farcs$ in diameter), and the ejecta ring shows substructures at several arcsec scales (Fig. 1). Therefore, we chose the ACIS-S3 data with an off-axis angle of $<2'$ as our primary data to fully utilize the high sensitivity in the soft X-ray band ($\it {E}$ $\leq$ 1 keV) and the superb spatial resolution ($\la 1\farcs$) of the ACIS. We reprocessed each individual ObsID with CIAO version 4.6 via the {\it chandra repro} script. We removed time intervals that show high particle background fluxes (a factor of $\sim2$ or more higher than the mean background flux level). After the data reduction, the total effective exposure time is $\sim 265$ ks. 

\begin{deluxetable}{cccc}[h]
\tabletypesize{\footnotesize}
\tablecaption{Observation log for archival {\it Chandra} ACIS-S3 data of E0102 used in this work}
\tablewidth{0pt}
\tablehead{\colhead{ObsID} & \colhead{Date} & \colhead{Exposure (ks)} & \colhead{Off-axis ($^\prime$)} } \\
\startdata

138\tablenotemark{a}	    & 1999 Aug 23   & 9.6	& 1.3	\\		
1231\tablenotemark{a}		& 1999 Aug 23   & 9.6	& 1.3	\\		
1423\tablenotemark{a}		& 1999 Nov 01   & 18.9	& 1.3	\\		
2850				& 2002 Jun 19  	 & 7.8	& 0.8	\\		
3544				& 2003 Aug 10   & 7.9   & 0.8	\\		
5130				& 2004 Apr 09	 & 19.4	& 0.7	\\		
6042				& 2005 Apr 12	 & 18.9	& 0.7	\\	
6043				& 2005 Apr 18	 & 7.9	& 1.3	\\		
6758				& 2006 Mar 19	 & 8.1	& 1.3	\\		
6759				& 2006 Mar 21	 & 17.9	& 0.7	\\
6765    			& 2006 Mar 19	 & 7.6	& 0.7	\\
6766				& 2006 Jun 06	 & 19.7	& 0.8	\\
8365				& 2007 Feb 11	 & 21.0	& 0.7	\\
9694				& 2008 Feb 07	 & 19.2	& 0.7	\\
10654				& 2009 Mar 01	 & 7.3	& 0.8	\\
10655				& 2009 Mar 01	 & 6.8	& 0.8	\\
10656				& 2009 Mar 06	 & 7.8	& 1.2	\\
11957				& 2009 Dec 30	 & 18.4	& 0.8	\\
13093				& 2012 Feb 01	 & 19.0	& 0.8	\\
14258				& 2012 Jan 12	 & 20.0	& 0.9   \enddata
\tablenotetext{a}{These ObsIDs were partially affected by moderately high particle background.}
\end{deluxetable}
\vspace{5mm}

\section{\label{sec:result} ANALYSIS \& RESULTS}
\subsection{X-Ray Images}

We created an ACIS-S3 broadband image, an X-ray three-color image, and a hardness ratio map of E0102 (Fig. 1). To produce these images we combined all individual observations. In three-color image red, green, and blue represent 0.4--0.8 keV, 0.8--1.16 keV, and 1.16--4 keV energy bands, respectively. We used the sub-band images with the native pixel scale of the ACIS detector ($0\farcs.49$ pixel$^{-1}$) and  adaptively smoothed them. In general, the main outermost boundary shows a nearly circular morphology. It also include very faint soft (red) X-ray emission (Fig. 1b). We created hardness ratio ({\it HR}) map following the methods by \cite{park06}. We define our hardness ratio as $HR=(H-S)/(H+S)$, where $H$ and $S$ refer to hard and soft energy bands, respectively \citep{park06}. In calculations of {\it HR}, we used 0.3--1.3 keV energy band for {\it S} and 1.3--3.0 keV energy band for {\it H}. While X-ray emission in the outermost boundary is generally soft, the center of SNR and outer parts of the bright ring show harder X-ray emission (Fig. 1c). The broadband intensity of the bright ring-like emission from the shocked ejecta peaks in brightness at $\sim14\farcs$--$15\farcs$ from the geometric center of the X-ray emission of the SNR (Fig. 1a). This intensity peak corresponds to red parts in the three-color image and soft X-ray emission regions in the hardness ratio map. Moreover, broadband gray-scale X-ray and {\it Hubble Space Telescope} (HST)\footnote{http://hst.esac.esa.int/ehst/\#home} optical images of E0102 for the purposes of comparison is shown in Fig. 2. While X-ray broadband image of E0102 shows a ring-like structure, the optical filaments are concentrate in the south, southeast, and southwest parts of the remnant. The bright ridge seen in the X-rays is evident also in the optical band. 

\begin{figure*}[htp]
\centerline{\includegraphics[angle=0,width=\textwidth]{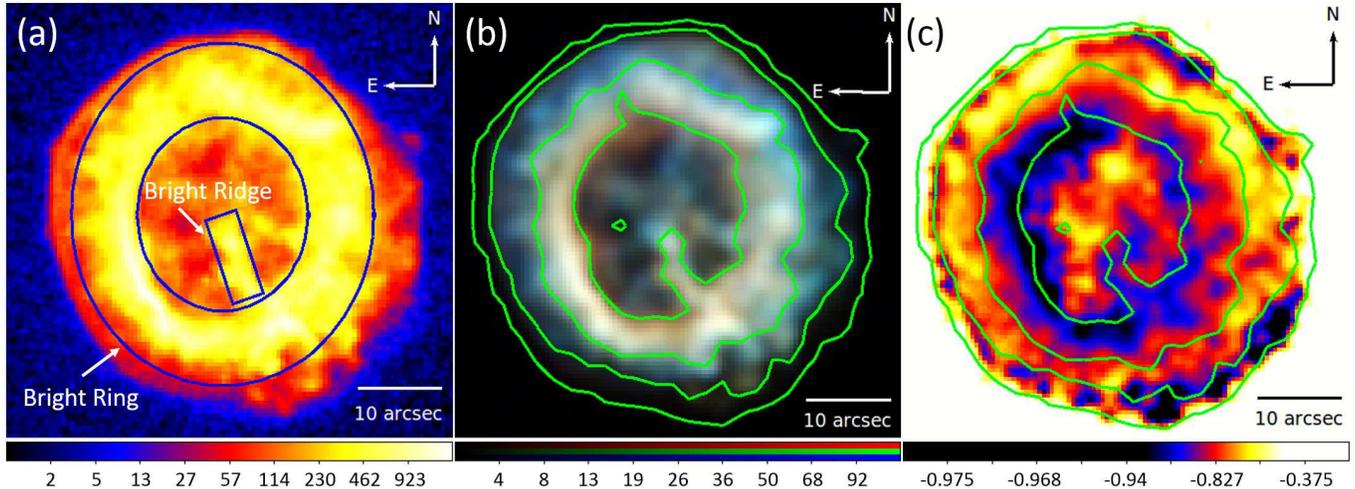}}
\figcaption[]{A log scale broadband image of E0102 (0.3--7 keV). It has been smoothed by a Gaussian kernel of $\sigma=0\farcs.25$. Bright ring and bright ridge regions are marked (a). A three-color image of E0102. The red is 0.1--0.8 keV, the green is 0.8--1.3 keV, and the blue is 1.3--5.0 keV band images. Each color image has been smoothed by a Gaussian kernel of $\sigma=0\farcs.75$ (b).  A hardness ratio map of E0102 ($HR=(H-S)/(H+S)$, $S$=0.3--1.3 keV, $H$=1.3--3.0 keV). The image has been smoothed by a Gaussian kernel of $\sigma=0\farcs.75$ for the purposes of display (c). All images are at the ACIS native pixel scale ($0\farcs.49$ pixel$^{-1}$). Broadband image contours of the SNR are overlaid on three-color and hardness ratio images.}\label{fig:fig1}
\end{figure*}

\begin{figure}[]
\centerline{\includegraphics[angle=0,width=1\columnwidth]{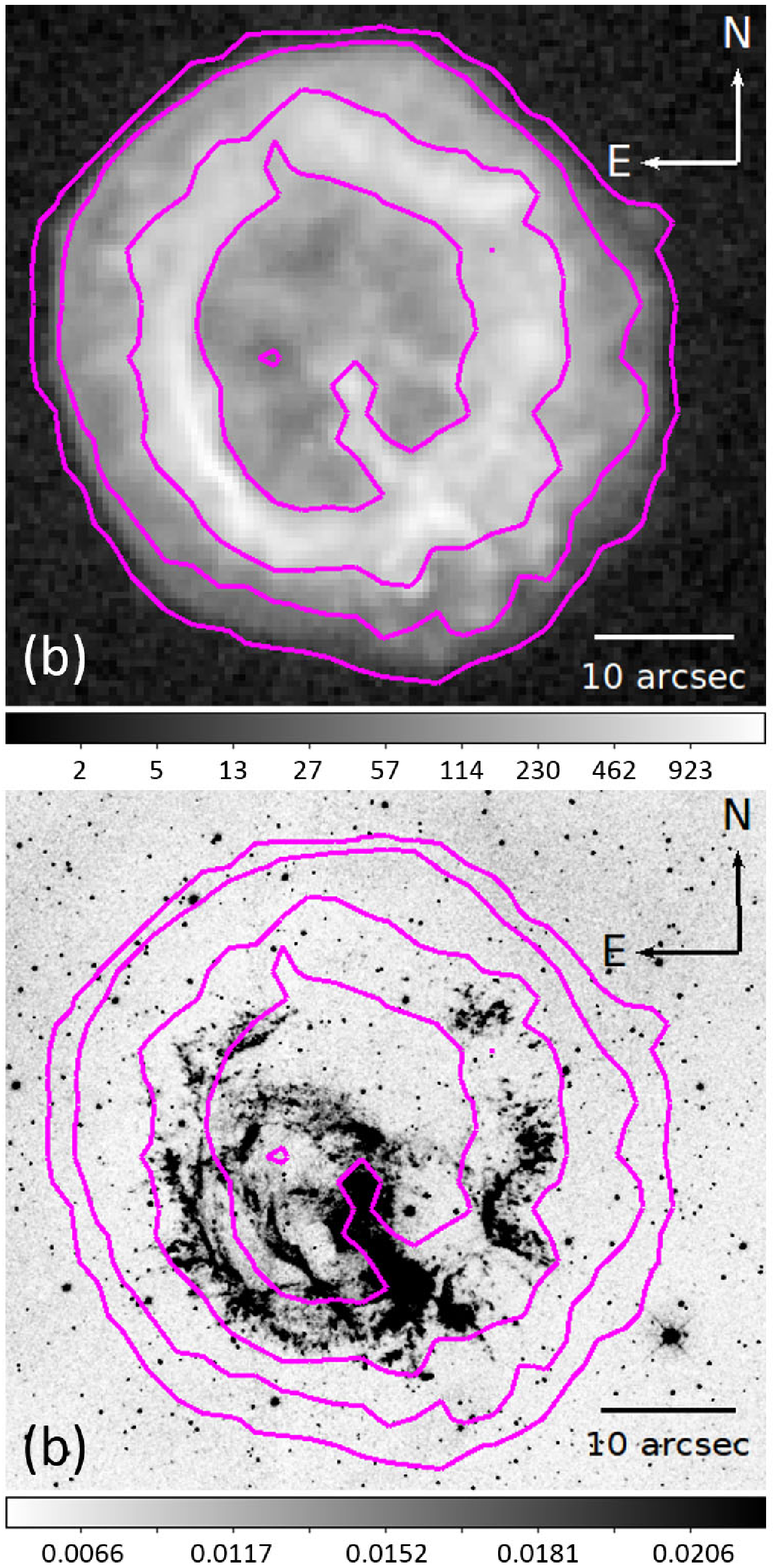}}
\figcaption[]{A logarithmic gray-scale broadband {\it Chandra} image of E0102 at the ACIS native pixel scale ($0\farcs.49$ pixel$^{-1}$). It has been smoothed by a Gaussian kernel of $\sigma=0\farcs.25$ for the purposes of display (a). A gray-scale HST optical image of E0102 at the WFC3/UVIS native pixel scale ($0\farcs.04$ pixel$^{-1}$. The image has been smoothed by a Gaussian kernel of $\sigma=0\farcs.75$ (b). Broadband image contours of the SNR are overlaid in each panel.}
\end{figure}

\subsection{Atomic Line Equivalent Width Images}

To investigate the distribution of atomic lines throughout E0102 we followed the method of \citet{hwang00} \citep[see also][]{park02} to construct equivalent width images (EWIs) for lines pronounced in the integrated spectrum (Fig. 3): O-He$\alpha$ ({\it E} $\sim$ 0.57 keV), O-Ly$\alpha$ ({\it E} $\sim$ 0.65 keV), Ne-He$\alpha$ ({\it E} $\sim$ 0.9 keV), Ne-Ly$\alpha$ ({\it E} $\sim$ 1.02 keV), Mg-He$\alpha$ ({\it E} $\sim$ 1.35 keV), and Si-He$\alpha$ ({\it E} $\sim$ 1.84 keV). To create these EWIs, we combined all 20 ObsIDs. The line bands used to create the EWIs are given in Table 2, along with the associated continuum bands. Before EW calculations, images in these bands were binned by 2$\times$2 pixels and then adaptively smoothed. The underlying continuum was calculated by logarithmically interpolating the fluxes from the higher and lower continuum images of narrow energy bands nearby the line to the line center energy. The estimated continuum flux was integrated over the line band and subtracted from the line emission flux. The continuum-subtracted line intensity was then divided by the estimated continuum flux on a pixel-by-pixel basis to generate the EWI for each line. The resulting EWIs are shown in Fig. 4. The EWIs serve primarily as a qualitative guide for an effective regional spectral analysis, and we made no attempt to interpret them quantitatively.

\begin{figure}[h]
\begin{flushleft}
\centerline{\includegraphics[trim=0.05cm 0.2cm 0cm 0.05cm, clip=true, width=1\columnwidth]{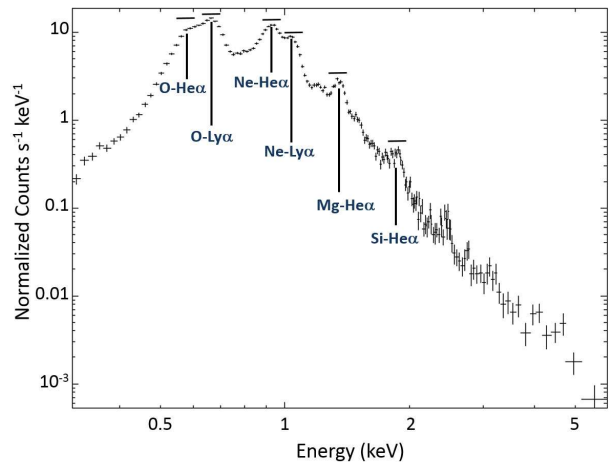}}
\figcaption[]{{An integrated spectrum of entire E0102 in the 0.3--7.0 keV band. Photon energy bands used for constructing the EWIs are marked.}\label{fig:fig2}}
\end{flushleft}
\end{figure}

\begin{deluxetable}{cccc}[htb]
\tabletypesize{\footnotesize}
\tablecaption{Photon Energy Bands Used for EWIs \label{table:ewibands}}
\tablewidth{0pt}
\tablehead{\colhead{Atomic Lines} & \colhead{Line} & \colhead{Low Continuum} & \colhead{High Continuum} \\
& (keV) & (keV) & (keV) }
\startdata
O He$\alpha$		& 0.51 -- 0.60	& 0.40 -- 0.50 & 0.74 -- 0.81 \\
O Ly$\alpha$		& 0.61 -- 0.74	& 0.40 -- 0.50 & 0.74 -- 0.81 \\
Ne He$\alpha$		& 0.84 -- 0.98	& 0.74 -- 0.81 & 1.12 -- 1.16 \\
Ne Ly$\alpha$		& 0.99 -- 1.10	& 0.74 -- 0.81 & 1.12 -- 1.16 \\
Mg He$\alpha$	    & 1.29 -- 1.42	& 1.25 -- 1.29 & 1.62 -- 1.70 \\
Si He$\alpha$	    & 1.75 -- 1.93	& 1.62 -- 1.70 & 2.02 -- 2.12
\enddata
\end{deluxetable}

\begin{figure*}[]
\label{section:ewi}
\centerline{\includegraphics[angle=0,width=\textwidth]{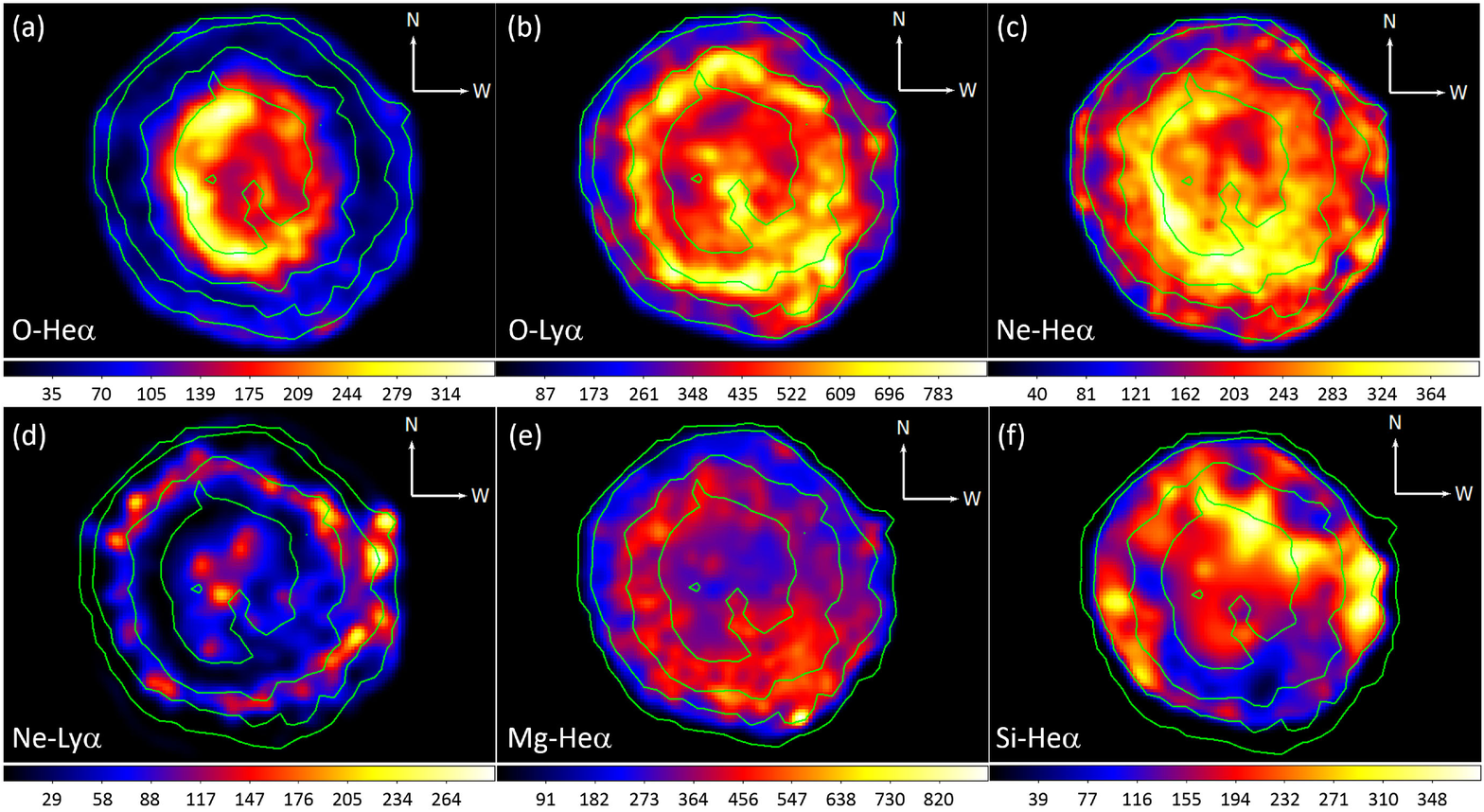}}
\figcaption[]{Linear scale line equivalent-width images for O-He$\alpha$ (a), O-Ly$\alpha$ (b), Ne-He$\alpha$ (c), Ne-Ly$\alpha$ (d), Mg-He$\alpha$ (e), and Si-He$\alpha$ (f) line emissions. Each EW image has been smoothed by a Gaussian kernel of $\sigma=0\farcs.75$. Broadband X-ray image contours of E0102 are overlaid.}
\end{figure*}

O-He$\alpha$ EW is enhanced along the projected interior of the bright ejecta ring (more intense in the east part than west part). O-Ly$\alpha$ EW is enhanced just outer of O-He$\alpha$ EW, and generally corresponds to the bright ring seen in the broadband image and it is more concentrated toward the outer boundary of the bright ring, especially in the southwest parts. That is, the lower-ionized O-He$\alpha$ line is enhanced near the inner boundary of the bright ring, and the higher-ionized O-Ly$\alpha$ line is enhanced in the outer boundary of the bright inner ring. These features are consistent with the progressive ionization of the O-rich ejecta gas shocked by the reverse shock propagating toward the geometric center of the X-ray emission of the SNR \citep{gaetz00}. O-Ly$\alpha$ EW is enhanced along the central bright ridge. Generally, O-He$\alpha$ and O-Ly$\alpha$ EWI maps correspond to the O-bright shocked filaments in optical HST image (Figs. 2 and 4). Ne-He$\alpha$ EW is enhanced in the inner boundary of the bright ring seen in the broadband image (specifically between the northeast and southwest parts). Ne-Ly$\alpha$ line is faint and appears to be enhanced just outside of the bright ring. Some enhanced Ne-Ly$\alpha$ line enhancements also seem to extend all the way out to the western outermost boundary of the SNR. Mg-He$\alpha$ EW is enhanced mainly in the east and south parts of the bright ring, while Si-He$\alpha$ EW is enhanced generally in the northern and western regions of the SNR and in the eastern rim. The outer shell beyond the bright inner ring generally shows no enhancement in any EWI except some O-Ly$\alpha$, Ne-He$\alpha$, and Ne-Ly$\alpha$ line enhancements outside of the ring in the west and southwest. Ne-Ly$\alpha$ shows small enhancements in the western outermost boundary and Ne-He$\alpha$ shows some enhancements beyond the bright ring in western shell. O-Ly$\alpha$ also shows small enhancements in the southwestern outermost boundary. Overall, the line EW distributions in E0102 are not symmetrical and enhanced emission lines show discrepancies in both radial and azimuthal directions. O, Ne and Mg line enhancements generally show {\bf a} ring like feature, but the Si EW distribution is far from such an organized ring-like structure.

\subsection{Spectroscopy}

Initially, we determined the X-ray spectrum from the outermost Shell region shown in Fig. 5a to determine characteristics of the swept-up CSM. Because of a more complicated mixture between emission features from the bright inner ejecta ring and the outer diffuse swept-up medium in the west and southwest regions of the SNR, these regions were excluded from our study of CSM. The Shell region (Fig. 5a) contains $\sim$11000 counts in the 0.3--7.0 keV energy band. In order to study spatial structure of X-ray emission from metal-rich ejecta gas, we chose 180 regions through nine radial directions across E0102 (Fig. 5b). Each radial direction is represented by the letters, from A to I, on the basis of the clockwise. The radial regions in these directions are also marked by numbers from the center of SNR to outward (Fig. 5b). Each region, shown in Fig. 5b contains $\sim$4500--5500 counts in the 0.3--7.0 keV energy band. We then extracted the spectra from each individual observation for all selected regions, and then combined them using the CIAO script {\it combine$\_$spectra}. We excluded ObsID138, ObsID1231, and ObsID1423, which are obtained in 1999, because of their significantly different quantum efficiency of the ACIS than the rest of the ObsIDs. The photon counts from these three ObsIDs contribute $\sim$12--15\% of the total flux in each regional spectrum, thus excluding these ObsIDs would not affect our results. We performed the background subtraction using the spectrum extracted from source-free regions outside of the SNR. We binned each spectrum to contain a minimum of 20 counts per energy channel. Our general approach for the spectral model fits for these regional spectra is as follows. We fit each regional spectrum with a plane-parallel shock model \citep{bork01} in non-equilibrium ionization (NEI) with two foreground absorption column components, one for the Galactic ($N_{\rm H,Gal}$) and the other for the SMC ($N_{\rm H,SMC}$). We used NEI version 2.0 (in xspec) associated with ATOMDB \citep{smith01}, which was augmented to include inner shell lines, and updated Fe-L lines \citep[see][]{bade06}. We fixed $N_{\rm H,Gal}$ at $4.5 \times 10^{20}$ cm$^{-2}$ for the direction toward E0102  \citep{DL90} with solar abundances \citep{ande89}. We fixed abundances in $N_{\rm H,SMC}$ at values by \cite{RD92}. We also fixed the redshift parameter at $z = 5.27 \times 10^{-4}$ for the radial velocity of the SMC \citep{richter87}.

\begin{figure*}[]
\centerline{\includegraphics[angle=0,width=\textwidth]{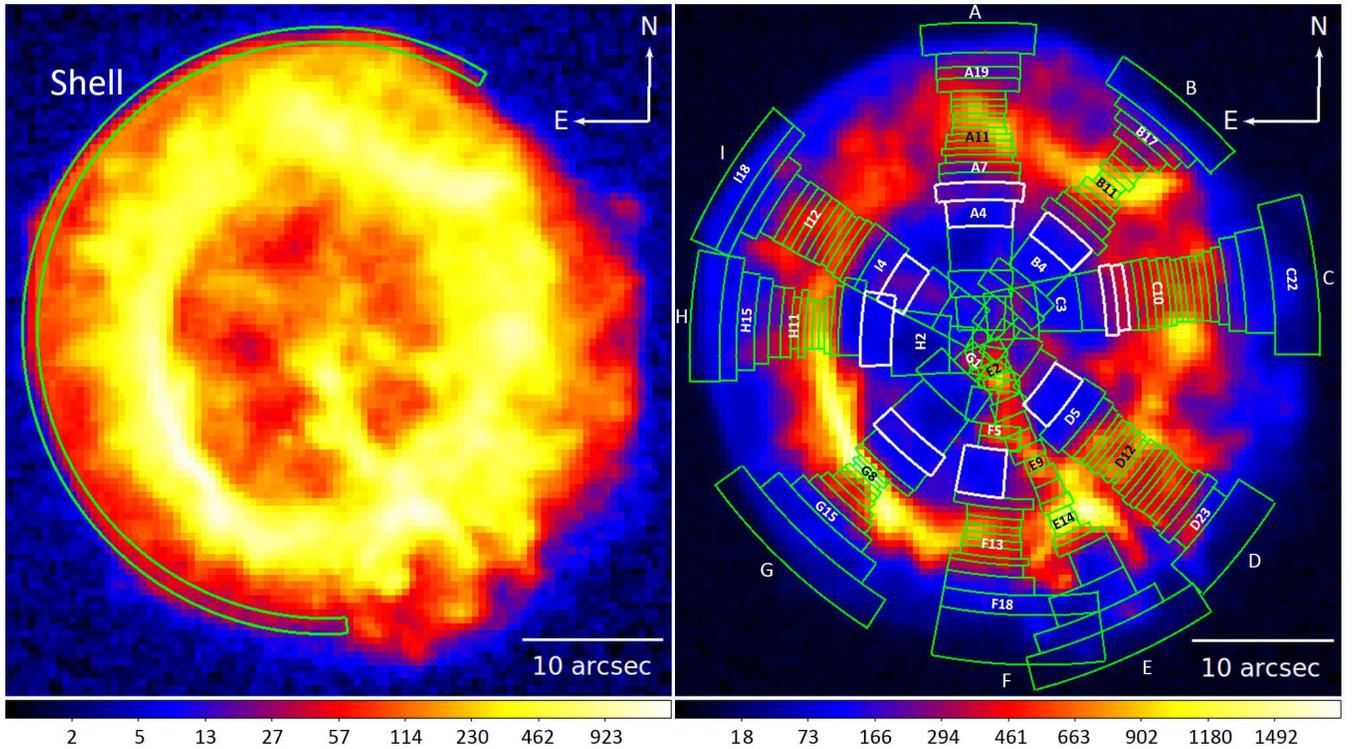}}
\figcaption[]{A log scale broadband image of E0102 in the 0.3--7 keV photon energy band. The outermost Shell region that used to characterize the spectral nature of CSM is marked. For the Shell region, the partial annulus thickness is $1\farcs.1$ and the outer radius is $21\farcs.6$ (a). A square-root scale broadband image of E0102 in the 0.3--7 keV photon energy band. Regions used for the spectral analysis are marked. White regions show \textquotedblleft{transition}" regions (see the text) (b). Both images are at the ACIS native pixel scale ($0\farcs.49$ pixel$^{-1}$), and they have been smoothed by a Gaussian kernel of $\sigma=0\farcs.25$.}\label{fig:fig5}
\end{figure*}

We fit the Shell region spectrum using a one-component plane-parallel shock model (Fig. 6). We initially fixed all elemental abundances at the SMC values, i.e. He = 0.832, C = 0.148, N = 0.038, O = 0.126, Ne = 0.151, Mg = 0.251, Si = 0.302, S = 0.240, Ar = 0.178, Ca = 0.214, Ni = 0.398, and Fe = 0.149 \citep{RD92}, in the plane-shock model. Abundances are with respect to solar values \citep{ande89} hereafter. We varied electron temperature ($kT$, where {\it k} is the Boltzmann constant), ionization timescale ($n_{e}t$, where $n_{e}$ is the post-shock electron density and $t$ is the time since the gas has been shocked) and $N_{\rm H,SMC}$ absorbing column in the SMC. The normalization parameter (a scaled volume emission measure, $\it EM$ = $n_en_HV$, where $n_H$ is the post-shock H density, and {\it V} is the emission volume) is also varied. The fit is shown in Fig. 6a was statistically rejected ($\chi_{\nu}^2 \sim 2.5$). Residuals from the model were significant around Ne, Mg and Si emission lines. We refit the data with Ne, Mg and Si abundances varied. This fit improved but it was still statistically unacceptable ($\chi_{\nu}^2 \sim 2.0$) as shown in Fig. 6b. We then varied O and Fe abundances to improve the fit. After we varied these parameters, the fit significantly improved and statistically acceptable ($\chi_{\nu}^2 \sim 1.0$) as shown in Fig. 6c. While the fitted O and Ne abundances are almost consistent with values given by \cite{RD92}, the fitted abundances for Mg, Si, and Fe significantly lower by a factor of $\sim 3 - 4$ than \cite{RD92} values. We conclude that the Shell region spectrum is dominated by emission from the shocked low-abundant CSM rather than that from the shocked metal-rich ejecta gas. The best-fit model parameters of the Shell region is summarized in Table 3. 

\begin{figure}[h]
\centerline{\includegraphics[angle=0,width=1\columnwidth]{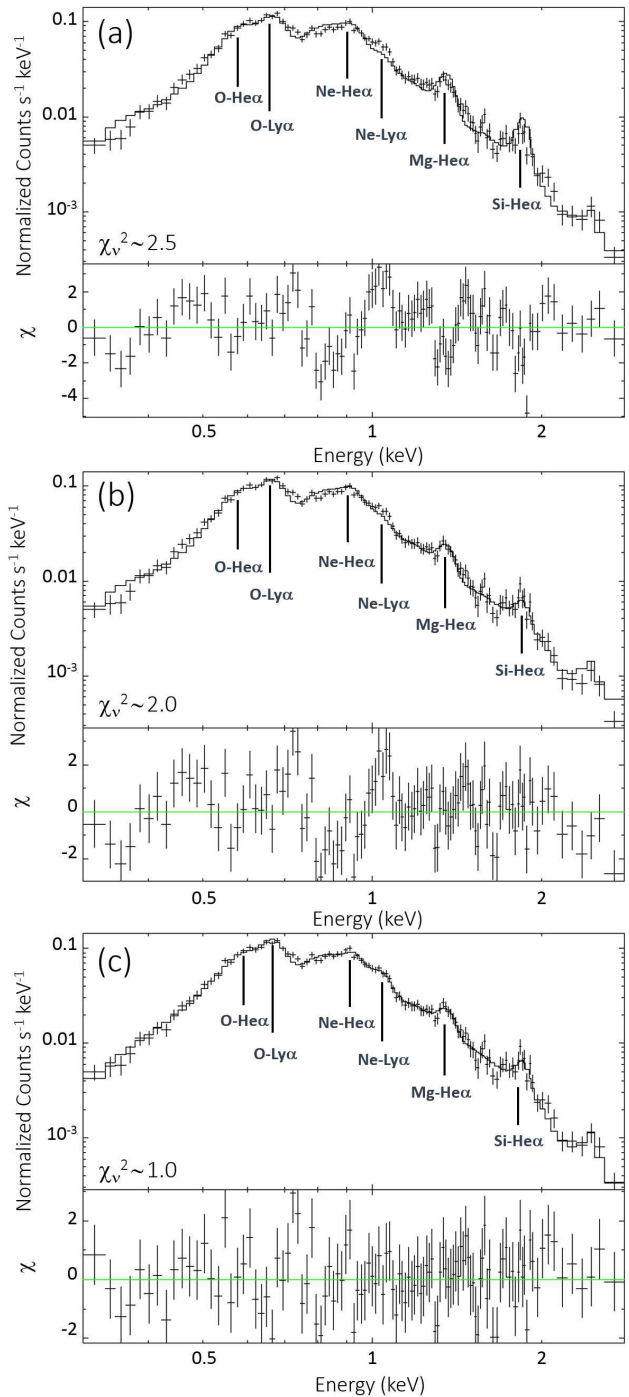}}
\figcaption[]{Spectral models and residuals of the Shell region. Several atomic emission line features are marked (a) All elemental abundances fixed SMC abundances \citep{RD92}. (b) Only Ne, Mg and Si abundances varied. (c) Best-fit spectral model.}
\end{figure}

\begin{deluxetable*}{ccccccccccc}
\footnotesize
\tabletypesize{\scriptsize}
\setlength{\tabcolsep}{0.05in}
\tablecaption{Summary of Spectral Model Fits to Shell Region of E0102}
\label{tbl:tab1}
\tablewidth{0pt}
\tablehead{\colhead{} & \colhead{$N_{\rm H,SMC}$} & \colhead{$kT$} & \colhead{$n_et$} & \colhead{\it EM} &\colhead{$\chi_{\nu}^2$} & \colhead{O} & \colhead{Ne} & \colhead{Mg} & \colhead{Si} & \colhead{Fe} \\
\colhead{} & \colhead{(10$^{21}$~cm$^{-2}$)} & \colhead{(keV)} & \colhead{(10$^{11}$ cm$^{-3}$ s)} & \colhead{($10^{57}$ cm$^{-3}$)} & \colhead{} & \colhead{} & \colhead{} & \colhead{} & \colhead{} & \colhead{}}
\startdata
Shell & $0.80^{+0.26}_{-0.35}$ & $0.58^{+0.10}_{-0.02}$ & $3.06^{+1.31}_{-0.88}$ & $20.62^{+5.29}_{-4.47}$ &  1.04 & $ 0.11^{+0.07}_{-0.02}$ & $0.13^{+0.04}_{-0.02}$ & $0.08^{+0.05}_{-0.01}$ & $0.13^{+0.01}_{-0.07}$ & $0.04^{+0.03}_{-0.01}$
\enddata
\tablecomments{Abundances are with respect to solar \citep{ande89}. Uncertainties are with a 90\% confidence level. The Galactic column $N_{\rm H,Gal}$ is fixed at 0.45 $\times$ 10$^{21}$ cm$^{-2}$. For comparisons the Russell \& Dopita (1992) values for SMC abundances are O = 0.126, Ne = 0.151, Mg = 0.251, Si = 0.302, and Fe = 0.149.}
\end{deluxetable*}

To investigate the detailed spatial and chemical structures of metal-rich ejecta in E0102, we perform an extensive spatially-resolved spectral analysis of its X-ray emission based on a number of small regional spectra (Fig. 5b). E0102 is a young supernova remnant and its spectrum is ejecta dominated \citep{gaetz00}. In fact, we found that the swept-up CSM emission component superposed on the ejecta-dominated bright regional spectra is small (the contribution from the Shell $\sim$5\% of the total flux). Thus, we fit our extracted regional spectra with one-component plane-shock model. Initially we varied $kT$, $n_{e}t$, and normalization while fixing $N_{\rm H,SMC}$, and all elemental abundances at the Shell values. The fits were statistically rejected for almost all regions ($\chi_{\nu}^2\ga4$). We then varied abundances for elements for which emission line features are associated with significant residuals from the fitted model. We found that it was necessary to vary O, Ne, Mg for all regions and Si for a few regions to obtain statistically acceptable fits ($\chi_{\nu}^2<1.6$). In this manner, we fitted 170 regional spectra and found out X-ray emission in subregions is dominated by emission from overabundant ejecta gas except northern and eastern outermost CSM boundary. We show some example spectra (see Figs. A1 and A2 in Appendix) extracted from regions marked in Fig. 5b and the best-fit spectral model with SMC abundances fixed model \citep{RD92} are overlaid in each panel. The results are based on our detailed spectral analysis in the northern (A), northwestern (B), western (C), southwestern (D), southern bright filament structure (E), southern (F), southeastern (G), eastern (H), and northeastern (I) parts of E0102 listed in Tables 5--13 in Appendix, respectively. Radial profiles of spectral parameters based on our analysis of regional spectra extracted in E0102 are shown in Figs. B1, B2 and B3 in Appendix. 

We had statistically poor fits ($\chi_{\nu}^2\ga2$) for 10 regional spectra with one-component plane-shock model. The poor fits are caused by the presence of both strong He$\alpha$ and Ly$\alpha$ lines for O and Ne in these regional spectra, which cannot be properly fitted by our simple single shock model. These regions are at a similar distance from the geometric center of the X-ray emission of the SNR as shown in Fig. 5b with white boxes named ``transition'' regions. In all other regions, He$\alpha$ and Ly$\alpha$ line structures from O and Ne ions are simpler (e.g. only O-He$\alpha$ and Ne-He$\alpha$ lines are enhanced, or only O lines are enhanced, etc), and that a single temperature model was able to fit the observed spectra. In contrast to that simple spectral structure, in these transition regions, both of He-$\alpha$ and Ly-$\alpha$ lines from both O and Ne ions are pronounced, for which the simple one-shock model cannot find an adequate solution (Fig. 7a). This means that in these regions a more complex spectral nature is implied: e.g., superpositions of emission from multiple component hot gas. Assuming a multiple component radiation, we attempted a two-component shock model to fit these spectra. For this scenario, we initially considered emission from the overabundant ejecta gas superposed on that from the surrounding low-abundant CSM shell. For the low-abundant component, presumably responsible for the superposed swept-up shell emission, we fixed all parameters at the Shell values except for normalization. The addition of the outer shell component did not improve the fits ($\chi_{\nu}^2\sim2$) because the ejecta component dominates the entire spectrum (the contribution from the low-abundant Shell component is only $\sim$5\% of the total flux). Therefore, we refit these spectra with a two overabundant ejecta components. We assumed that two ejecta components have the same abundances, but different electron temperatures, ionization states and emission measures. Thus, we tied these elemental abundances for two different shock components to each other, and we thawed O, Ne and Mg abundances. In this way, we found statistically acceptable fits ($\chi_{\nu}^2\sim1$) for these ten regions. As an example, we show our best-fit two-component ejecta spectral model fit for the region A4 spectrum in Fig. 7b. Furthermore, there are two highly overabundant regions, G1 and H1, near the central part of E0102 in projection. Calculated elemental abundances of these regions are $\sim 30-70$ times higher than CSM values. Spectral models of G1 and H1 regions are also shown in Fig. 8. 

\begin{figure}[htp]
\centerline{\includegraphics[angle=0,width=1\columnwidth]{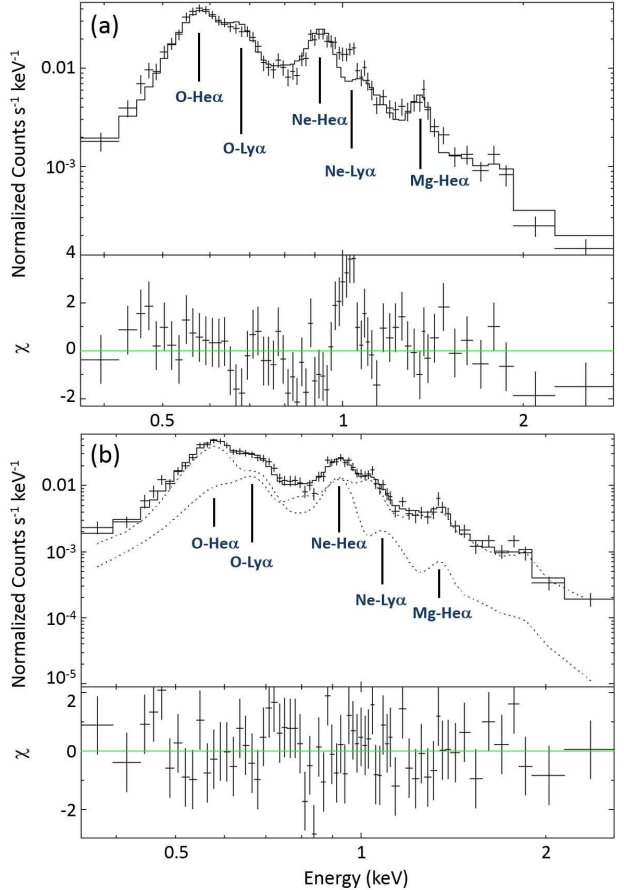}}
\figcaption[]{The single shock (a) and two shock (b) spectral models and residuals of the A4 region. Several atomic emission line features are marked.}
\end{figure}

\section{\label{sec:disc} DISCUSSION}
\subsection{CSM Structure and Outermost Boundary Feature of E0102}

We examined the X-ray spectrum from outermost Shell region to determine characteristics of the swept-up CSM. Our results show abundances of Mg, Si and Fe are markedly lower (by a factor of $\sim 3 - 4$) than those by \cite{RD92}. This suggests E0102 might have exploded in a locally metal-poor medium, which may support a low-$Z$ progenitor \citep[][]{schenck14}, or the heavy elements simply might be depleted onto dust grains \citep{kozasa91, marassi18}. \citet{stanimirovic05} showed that the forward shock does not necessarily destroy all dust grains for E0102. This supports that some heavy elements may be locked in the dust grains. We estimate electron temperature and ionization timescale of the Shell as 0.58 keV and $3.06\times10^{11}$ cm$^{-3}$ s, respectively. We also measured column density for the Shell as $0.80\times10^{21}$ cm$^{-2}$.  

\begin{deluxetable*}{cccccccc}[p]
\footnotesize
\tabletypesize{\scriptsize}
\setlength{\tabcolsep}{0.05in}
\tablecaption{A Comparison between \cite{hughes00} and this work}
\label{tbl:tab4}
\tablewidth{0pt}
\tablehead{ \colhead{} & \multicolumn{2}{c}{\cite{hughes00}} & \colhead{} & \colhead{} & \colhead{This Work} &\colhead{} & \colhead{}  \\
\colhead{Parameters} & \colhead{Single Ionization Timescale} & \colhead{Planar Shock} & \colhead{Shell} & \colhead{A21} &\colhead{G17} & \colhead{H17} & \colhead{I18}} \\
\startdata
{$kT$ (keV)} & {$0.48^{+0.08}_{-0.05}$} & {$0.78^{+0.16}_{-0.15}$} & {$0.58^{+0.10}_{-0.02}$} & {$0.49^{+0.05}_{-0.04}$} & {$0.57^{+0.06}_{-0.05}$} &{$0.55^{+0.05}_{-0.06}$} & {$0.56^{+0.04}_{-0.04}$} \\
{$n_et$ (10$^{11}$ cm$^{-3}$ s)} & {$4^{+21}_{-2}$} & {$4^{+4}_{-2}$} & {$3.06^{+1.31}_{-0.88}$} & {$2.26^{+11.45}_{-5.63}$} & {$4.32^{+3.83}_{-1.87}$} & {$6.60^{+5.09}_{-2.91}$} & {$7.27^{+2.84}_{-2.90}$} \\
{$N_{\rm H,SMC}$ (10$^{21}$~cm$^{-2}$)} & {$0.5^{+0.3}_{-0.3}$} & {$<0.25$} & {$0.80^{+0.26}_{-0.35}$} & {$0.80\tablenotemark{a}$} & {$0.80\tablenotemark{a}$} & {$0.80\tablenotemark{a}$} & {$0.80\tablenotemark{a}$} \\
{$n_{e}n_{H}V$ /$4 \pi D^{2}$ (10$^{10}$ cm$^{-5}$)} & {$11^{+4}_{-3}$} & {$4.1^{+2.0}_{-0.8}$} & {$4.78^{+1.23}_{-1.04}$} & {$1.81^{+0.22}_{-0.28}$} & {$1.33^{+0.21}_{-0.20}$} & {$1.82^{+0.32}_{-0.21}$} & {$1.58^{+0.21}_{-0.19}$} \\
{O} & {$0.26^{+0.19}_{-0.14}$} & {$0.34^{+0.15}_{-0.12}$} & {$0.11^{+0.07}_{-0.02}$} & {$0.22^{+0.08}_{-0.07}$} & {$0.14^{+0.08}_{-0.04}$} & {$0.14^{+0.07}_{-0.04}$} & {$0.17^{+0.07}_{-0.05}$} \\
{Ne} & {$0.31^{+0.11}_{-0.09}$} & {$0.67^{+0.20}_{-0.19}$} & {$0.13^{+0.04}_{-0.02}$} & {$0.20^{+0.06}_{-0.05}$} & {$0.18^{+0.06}_{-0.04}$} & {$0.18^{+0.06}_{-0.04}$} & {$0.22^{+0.06}_{-0.05}$} \\
{Mg} & {$0.28^{+0.12}_{-0.10}$} & {$0.47^{+0.16}_{-0.16}$} & {$0.08^{+0.05}_{-0.01}$} & {$0.11^{+0.05}_{-0.04}$} & {$0.10^{+0.04}_{-0.04}$} & {$0.07^{+0.04}_{-0.03}$} & { $0.13^{+0.05}_{-0.04}$} \\
{Si} & {$0.302\tablenotemark{b}$} & {$0.302\tablenotemark{b}$} & {$0.13^{+0.01}_{-0.07}$} & {$0.13\tablenotemark{a}$} & {$0.13\tablenotemark{a}$} & {$0.28^{+0.13}_{-0.10}$} & {$0.13\tablenotemark{a}$} \\
{Fe} & {$0.010^{+0.005}_{-0.005}$} & {$0.06^{+0.03}_{-0.03}$} & {$0.04^{+0.03}_{-0.01}$} & {$0.04\tablenotemark{a}$} & {$0.04\tablenotemark{a}$} & {$0.04\tablenotemark{a}$} & {$0.04\tablenotemark{a}$} \\
{$\chi_{\nu}^2$} & {1.50} & {1.21} & {$1.04$} & {1.05} & {1.31} & {1.09} & {1.11}
\enddata
\tablecomments{Abundances are with respect to solar \citep{ande89}. For comparisons the \cite{RD92} values for SMC abundances are O = 0.126, Ne = 0.151, Mg = 0.251, Si = 0.302, and Fe = 0.149.}
\tablenotetext{a}{Parameter was fixed at the best-fit Shell value}
\tablenotetext{b}{Si was fixed at \cite{RD92} value}
\end{deluxetable*}

We evaluate the outermost region in each radial direction to reveal azimuthal structure of the CSM. The elemental abundances and ionization timescales are an order of magnitude higher than the Shell values in the western outermost boundary, and electron temperature a little higher (0.72 keV) than the Shell (0.58 keV). In addition, abundances are relatively higher and the ionization timescales significantly higher by a factor of $\sim 5-10$ than the Shell values in the southwestern outermost boundary. These may suggest an asymmetric explosion or possibly a density gradient in the circumstellar medium (mass loss from the progenitor star). Otherwise, the electron temperatures are consistent ($\sim \langle  0.58\rangle$ keV) azimuthally in the outermost boundaries for all directions.

We compare our results for the Shell region with the values given by \cite{hughes00} who examined the similar outer region of E0102 (Table 4). For these comparisons, we also used the results from our several regional spectral analysis (regions A21, G17, H17, and I18). The O, Ne, Mg and Si abundances calculated in our study were 2-5 times lower than the \cite{hughes00} values, while the Fe abundances are consistent with each other. Our data (a total of $\sim 235$ ks exposure of the {\it Chandra} ACIS) are more than an order of magnitude deeper than those used by \cite{hughes00} (8.8 ks exposure of the {\it Chandra} ACIS), providing a significantly higher $S/N$. These improved photon count statistics allow us more accurate estimates of elemental abundances. We note that, based on deep 450 ks {\it Chandra} ACIS observations of SNR B0049-73.6, \cite{schenck14} measured  similarly low SMC abundances to our values. \cite{hughes00} fixed their Si abundance at \cite{RD92} value (0.302) which is nearly 3 times larger than our measurement. This may have also contributed the overall higher abundance values estimated by \cite{hughes00}. These elemental abundance measurements are also model-dependent (as shown in Table 4). Our model utilizes improved atomic data than those \cite{hughes00} used. Furthermore, thanks to the significantly higher photon count statistics, we are able to extract the Shell spectrum from a thinner ($1\farcs.1$ in thickness, compared to $1\farcs.5$ in Hughes et al. 2000) and outer region ($\sim 22\farcs$ from the geometric center of the SNR, compared to $21\farcs$ in Hughes et al. 2000). This detail in the outer shell region selection may also partially contribute the differences in the elemental abundance measurements between two works. In fact, in Table 4, regions A21, G17, H17, and I18 are thicker than the Shell region, and show slightly higher abundances (although errors are overlapping with the Shell values).

We do not find any non-thermal synchrotron filaments along the outermost boundary, although strong cosmic-ray (CR) acceleration has been indicated in this SNR by \cite{hughes00}. Currently there is no signature for CR acceleration.

\begin{figure}
\centerline{\includegraphics[angle=0,width=1\columnwidth]{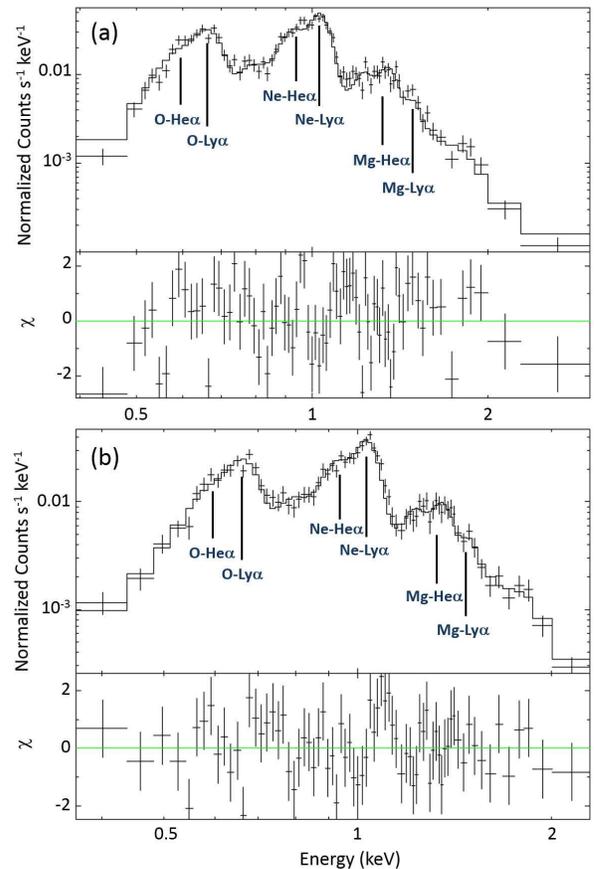}}
\figcaption[]{Best-fit models and residuals of X-ray spectrum of the highly overabundant G1 (top panel) and H1 (bottom panel) regions. Several atomic emission line features are marked.}
\end{figure}

\subsection{Spatial and Chemical Structures of Ejecta}

The ionization timescale of the metal-rich ejecta gas in E0102 generally peaks at $\sim 15\farcs - 19\farcs$ from the SNR's X-ray geometric center and decreases down to radius $r\sim 10\farcs$ for all radial directions. The ionization timescale also decreases beyond $r\sim 20\farcs$. The area from $r \sim 10\farcs$ to $19\farcs$ generally coincident with the bright ejecta ring (see green lines in Figs. B1, B2, and B3 in Appendix). This ionization structure supports the progressive ionization of the reverse-shocked ejecta as suggested by radial profiles of  emission line fluxes \citep{gaetz00}. Elemental abundances in most directions are low (around CSM abundances) near the SNR's outer boundary. The abundances rise sharply from $r \sim 19\farcs$ until $r\sim 15\farcs-16\farcs$, by a factor of $\sim 10-15$ of CSM values, then decrease somewhat but stay enhanced (by a factor of $\sim 3-4$ higher than the CSM abundances). These overall radial abundance profiles suggest that the contact discontinuity (the boundary between CSM and ejecta) is located at $r\sim 19\farcs$, generally corresponding the outer boundary of the bright ring. Assuming that emission from the reverse shocked overabundant gas may be emphasized along the lines of sight through the ejecta shell behind the reverse shock front, we conclude the current reverse shock position at $r\sim 15 \farcs -16\farcs$ from the geometric center of the X-ray emission of the SNR. The emission measure peaks at $r\sim 14\farcs-15\farcs$, conforming to the intensity peak of the ejecta ring, and it shows a sharp drop toward the SNR center. It is difficult to interpret these radial structures of spectral parameters in terms of the standard core-envelope models. An analytic solution \citep{chevalier82}, which describes the interaction of power law ejecta envelope with circumstellar winds with a density profile of $\rho \propto r^{-2}$, shows that both the shocked CSM and ejecta gas have the density peak at the contact discontinuity. Such a behavior is also reproduced by numerical simulations \citep{dwarkadas05, dwarkadas07}. 

Complex structures in the CSM could have affected the structure of shocked ejecta. For instance, numerical simulations show that, when the outer blast wave encounters a dense wind shell, a reflected shock is developed \citep[along with the transmitted shock into the shell, e.g.,][]{dwarkadas05}. Such a shock propagating back to the geometric center of the X-ray emission of the SNR might produce inwardly increasing density structure. The transmitted shock into the dense shell will enhance the X-ray emissivity of the shocked CSM. The limb-brightening in the northeastern boundary (see Fig. 1a), in contrast to the faint and diffuse outer boundary in southwest, might suggest such an interaction between the blast wave and the dense wind shell. Alternatively, the non-standard structure of the shocked ejecta might have been caused by the inherent complexity in the ejecta, such as a ring-like structure suggested by \cite{hughes94} or a Fe-Ni bubble. The \textquotedblleft{bubble}" effect may produce a low density region in the interior of ejecta \citep{chevalier05} and a ring-like geometry of ejecta can occur as a result of the core-collapse process, through the interaction of the CSM or a mix of the two \citep{blodin96, khokhlov99, wang01}. Our spectral analysis indicates that the innermost regions of the ejecta ring are hotter with lower ionization timescales than the outer regions (see Fig. 9 and Figs. B1, B2, and B3 in Appendix). This may suggest that the reverse shock is propagating into a low-density region near the geometric center of the X-ray emission of the SNR, which could be the projected low-density gas surrounding the ejecta ring, or could be the interior of the Fe-Ni bubble.

\begin{figure*}[]
\centerline{\includegraphics[angle=0,width=\textwidth]{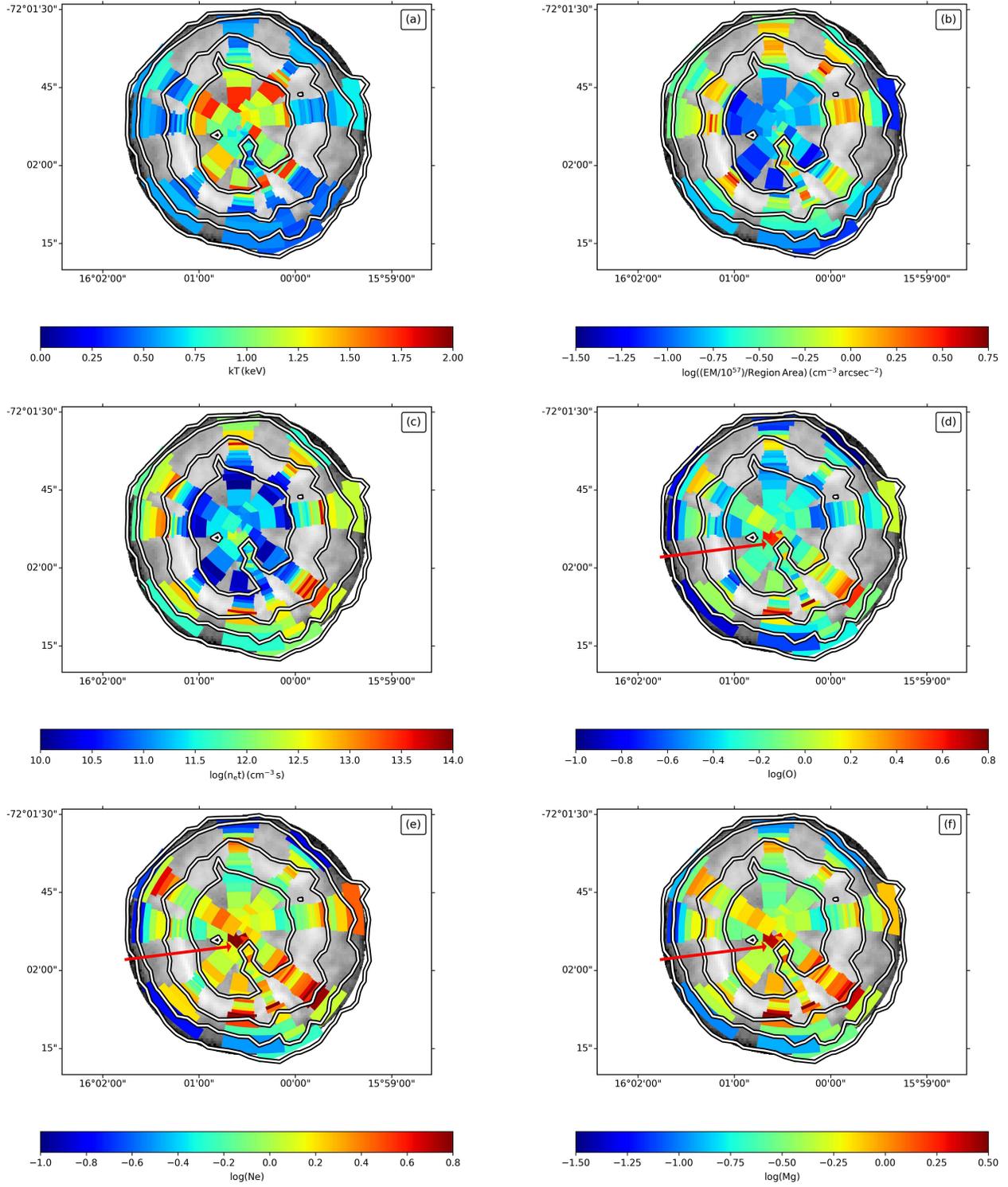}}
\figcaption[]{{Electron temperatures; $kT$ (a), emission measures; $EM$ (b), ionization timescales; $n_{e}t$ (c), O abundance (d), Ne abundance (e), and Mg abundance (f) distribution on E0102. Broadband image contours of E0102 are overlaid. The red arrow on the abundance plots points to the highly overabundant region around the center of the SNR. Abundances are with respect to solar \citep{ande89} and given logarithmically. Average plasma parameters ($kT$, $n_{e}t$, $EM$) are given for transition regions that applied two shock model.}\label{fig:fig10}}
\end{figure*}

Elemental abundances are not steady along radius. They generally slope down from the geometric center of the X-ray emission of the SNR until $r\sim 10\farcs-11\farcs$, then increase and reach maximum at  $r\sim 15\farcs-16\farcs$. After this, abundances decrease again and reach CSM values at $r\sim 19\farcs-20 \farcs$. This structure of elemental abundances is generally inconsistent with the projected X-ray emission from a shocked spherical shell of ejecta, as suggested by previous studies \citep[e.g.,][]{hughes94, flanagan04}. 

Distributions of the elemental abundances in the western and southwestern parts of the SNR are significantly different from that in other directions. In the west direction, elemental abundances decrease until $r \sim 9\farcs-10\farcs$, later they ascend and peak at $r\sim16\farcs-17\farcs$, and then decrease until $r\sim 19\farcs$. After $r\sim 19\farcs$, elemental abundances increase again. In contrast to other directions, the elemental abundances show an order of magnitude higher values than CSM in the western outermost boundary. Thus, metal-rich ejecta gas appears to extend further outward in west. There are two bases at $r\sim 5\farcs$ and $r\sim 14\farcs$, and two peaks at $r\sim 11\farcs$ and $r\sim 17\farcs$ for elemental abundances in the southwest direction. Moreover, values of elemental abundances still higher than CSM at the outermost boundary of this direction.  Also, in the south and southwestern directions, elemental abundances are almost two times higher than others (Fig. 9 and Figs. B1, B2, and B3 in Appendix). These complex structures of the elemental abundances between the western and southern parts of ejecta may support an asymmetric explosion as suggested by \cite{vogt10}.

Elemental abundances in the bright ridge connecting the bright ring and SNR's center are relatively steady compared to the other directions and consistently enhanced, by a factor of $\sim$10--20 higher than CSM values, throughout the radius. This indicates that this bright southwestern ridge is ejecta-dominated emission. 

Our spectral analyses indicate that there is a highly overabundant region, by a factor of $\sim 30-70$ of CSM values, in the central part of the SNR (see Fig. 9). Being consistent with the EWI maps, there are strong O-Ly$\alpha$ and Ne-Ly$\alpha$ emission lines in this region (Fig. 8). Also, both of Mg-Ly$\alpha$ and He$\alpha$ lines are enhanced in these central regions. This is in contrast to the rest of the SNR where Mg-Ly$\alpha$ lines are generally not enhanced. The spectrally hard region in the southeast of the geometric center of the X-ray emission of the SNR (as shown in Fig. 1c) and the blue emission on the southeast side of the central tip of the ejecta ridge in the three-color image are spatially coincident with this highly overabundant region (see Fig. 1b). Furthermore, this region corresponds to the blueshifted region ($-1800$ km s$^{-1}$), which was identified by the high resolution X-ray spectroscopic study \citep{flanagan04}. This shows that there is clumpy overabundant ejecta gas in the near of the geometric center of the X-ray emission of the SNR in projection.

We for the first time measure the projected spatial distributions of elemental abundances (for O, Ne, Mg, and Si), electron temperature, ionization timescale, and gas density in E0102. Based on these results, we find that metal-rich ejecta in E0102 clearly show asymmetric distributions. The apparently ring-like overall ejecta morphology in this SNR appears to be a projection of a more complex 3-D structure shocked gas rather than a 2-D projection of a simple spherical shell-like hot gas.

\subsection{Transition Regions}

X-ray spectra extracted from \textquotedblleft{transition}" regions (as marked in Fig 5b) cannot be fitted with one-component shock model. These regions are projected near the inner boundary of the bright ejecta ring at about $9\farcs$ from the center of the SNR. Spectra of these regions have both strong He$\alpha$ and Ly$\alpha$ lines for O and Ne and a single temperature and ionization state model cannot fit all of those lines. This suggests that the observed X-ray emission may be a superposition of those from multiple gas components with characteristically different thermal states. This structure appears in all directions except bright ridge in the southwest. Our spectral model fits to these regional spectra show that emission two distinctive gas components are required: i.e., one with a higher electron temperature ($kT$ $\sim 1.3$ keV) and lower ionization timescale ($n_{e}t\sim2\times10^{10}$ cm$^{-3}$ s) and the other with a lower temperature ($kT$ $\sim 0.5$ keV) and a higher ionization timescale ($n_{e}t \sim3.8\times10^{12}$ cm$^{-3}$ s). The hot component is generally consistent with emission from the shocked ejecta gas prevailing in the SNR. The cooler emission component shows a very high ionization state and a low electron temperature differently from the general distribution of ejecta gas parameters. It is difficult to interpret the spectral nature of this transition regions in the context of a simple spherical shell picture. \\

Our spatially-resolved spectral analysis results suggest that morphology of E0102 is not simple spherical shell. The overall ejecta features show that the spatial distribution of metal-rich stellar debris in E0102 is far from spherically symmetric. While the outermost forward shock boundary of E0102 is nearly circular, the metal-rich ejecta has an asymmetric distribution. The nature of explosion may be asymmetric and ring like structure of the remnant may originate its nature as suggested by \cite{flanagan04}, \cite{finkelstein06}, and \cite{vogt10}. Although the general structure of E0102 looks like cylindrical or toroidal geometry suggested by \cite{flanagan04}, the overall geometry of the SNR is likely more complicated. A more complicated 3-D model would be required to reveal the true spatial structure of E0102, which is beyond the scope of this work.

\subsection{Progenitor Nature and SNR Dynamics}
Based on the average values of the ejecta elemental abundances measured from individual ejecta regions, we estimate the abundance ratios of O/Ne $=3.6^{+2.4}_{-1.3}$, O/Mg $=29.4^{+20.5}_{-12.2}$, and Ne/Mg $=8.2^{+5.4}_{-3.2}$. These abundance ratios are in plausible agreement with the core-collapse supernova nucleosynthesis models for a 40$M_{\odot}$ progenitor with solar or sub-solar ({\it Z} = 0.004) metallicity \citep{nomo06}. This progenitor mass is higher than that estimated by \cite{flanagan04} (32 $M_\odot$), and lower than that estimated by \cite{finkelstein06} (50 $M_\odot$). The Shell is significantly underabundant compared to the general SMC values \citep{RD92} for Mg, Si and Fe. This suggests that E0102 might have exploded in a locally metal-poor environment.

Based on the best-fit volume emission measure ({\it EM}) in our spectral model fits of the Shell we estimate the post-shock electron density ({\it n$_e$}). For this estimation we assumed the X-ray emitting volume of {\it V} $\sim6.85 \times 10^{55}$ cm$^3$ for the Shell region for a $\sim$0.4 pc path length (roughly corresponding to the angular thickness of the Shell at 60 kpc) along the line of sight. We assumed $n_e\sim1.2 n_H$ for a mean charge state with normal composition (where $n_H$ is the H number density). Under the assumption of electron--ion temperature equipartition for E0102, the gas temperature is related with the shock velocity ($v_s$) as $T=3\hat{m}v_s^2/16k$ (where $\hat{m}\sim0.6m_p$ and $m_p$ is the proton mass). For gas temperature of $kT=0.6$ keV for the outer swept-up shell, we calculate a shock velocity of $v_s$ $\sim$ 710 km s$^{-1}$. For the SNR radius of $\sim$6.4 pc we estimate the Sedov age of $\tau_{sed} \sim3500$ yr for E0102. Our shock velocity estimate is only a conservative lower limit and therefore our SNR age estimate would be an upper limit. Although our age upper limit is not tightly constraining, it is generally consistent with previous estimates of $\sim1000-2000$ yr \citep{tuohy83, hughes00, eriksen01, finkelstein06}. We also calculate the corresponding explosion energy of $E_0\sim1.8 \times 10^{51}$ erg. \\

\acknowledgments
We thank the anonymous referee for his/her insightful and constructive suggestions, which significantly improved the paper. This work has been supported in part by {\it Chandra} grant AR3-14004X, the Scientific and Technological Research Council of Turkey (T\"UB\.ITAK)[BIDEB-2214, BIDEB-2211 Scholarship Programs], and  Scientific Research Project Coordination Unit of Istanbul University, project number: 23292. We thank Jayant Bhalerao, Andrew Schenck and Seth A. Post for their help in the Chandra data analysis. We also would like to thank Olcay Plevne for his contributions to the creation of the figures.

\appendix

\begin{figure*}[]
\figurenum{A1}
\centerline{\includegraphics[angle=0,width=\textwidth]{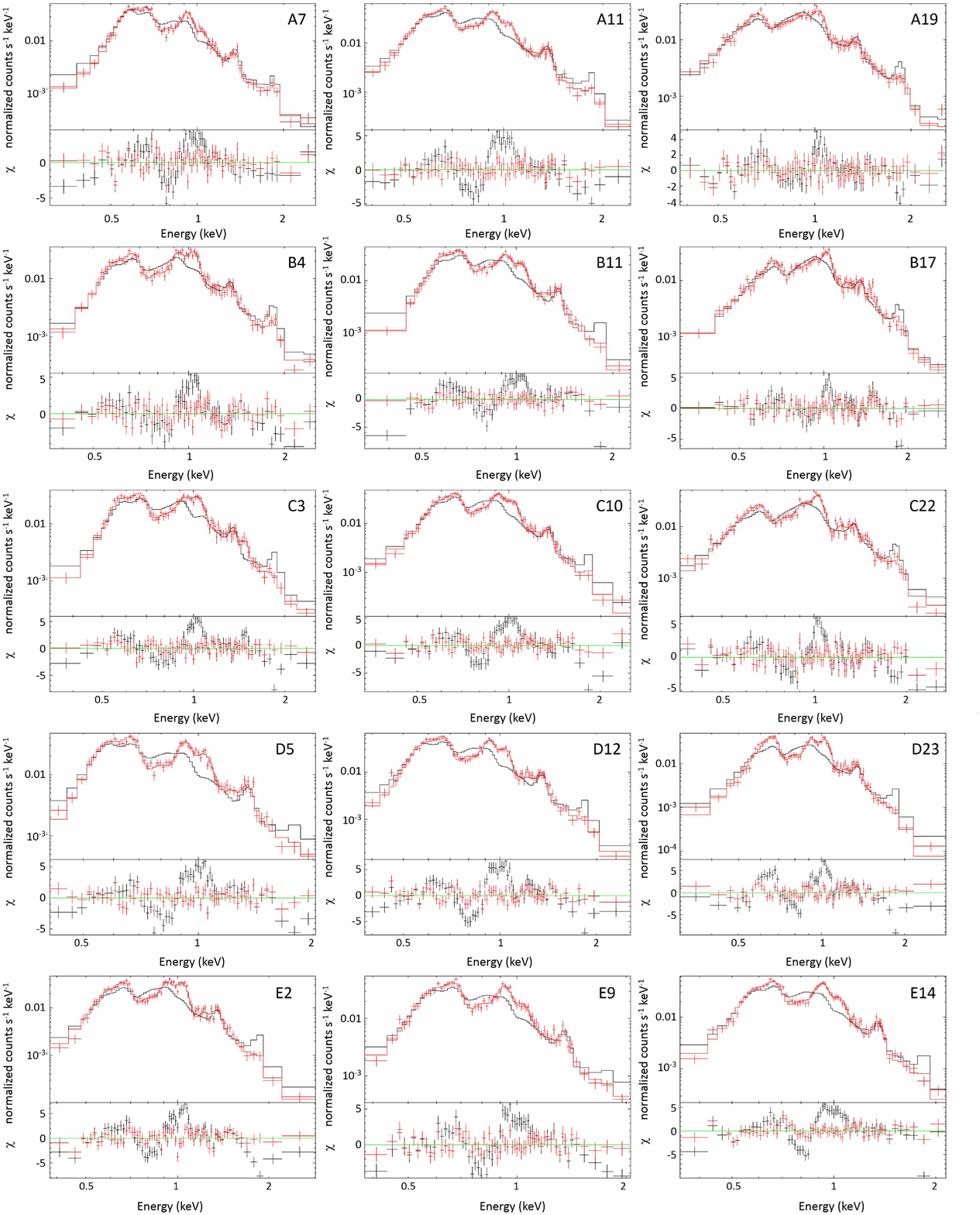}}
\figcaption[]{A small sample of the observed X-ray spectra extracted from regions shown in Fig. 5b. The best-fit one-component shock model with SMC abundances fixed model \citep{RD92} are overlaid in each panel.}
\end{figure*}

\begin{figure*}[]
\figurenum{A2}
\centerline{\includegraphics[angle=0,width=\textwidth]{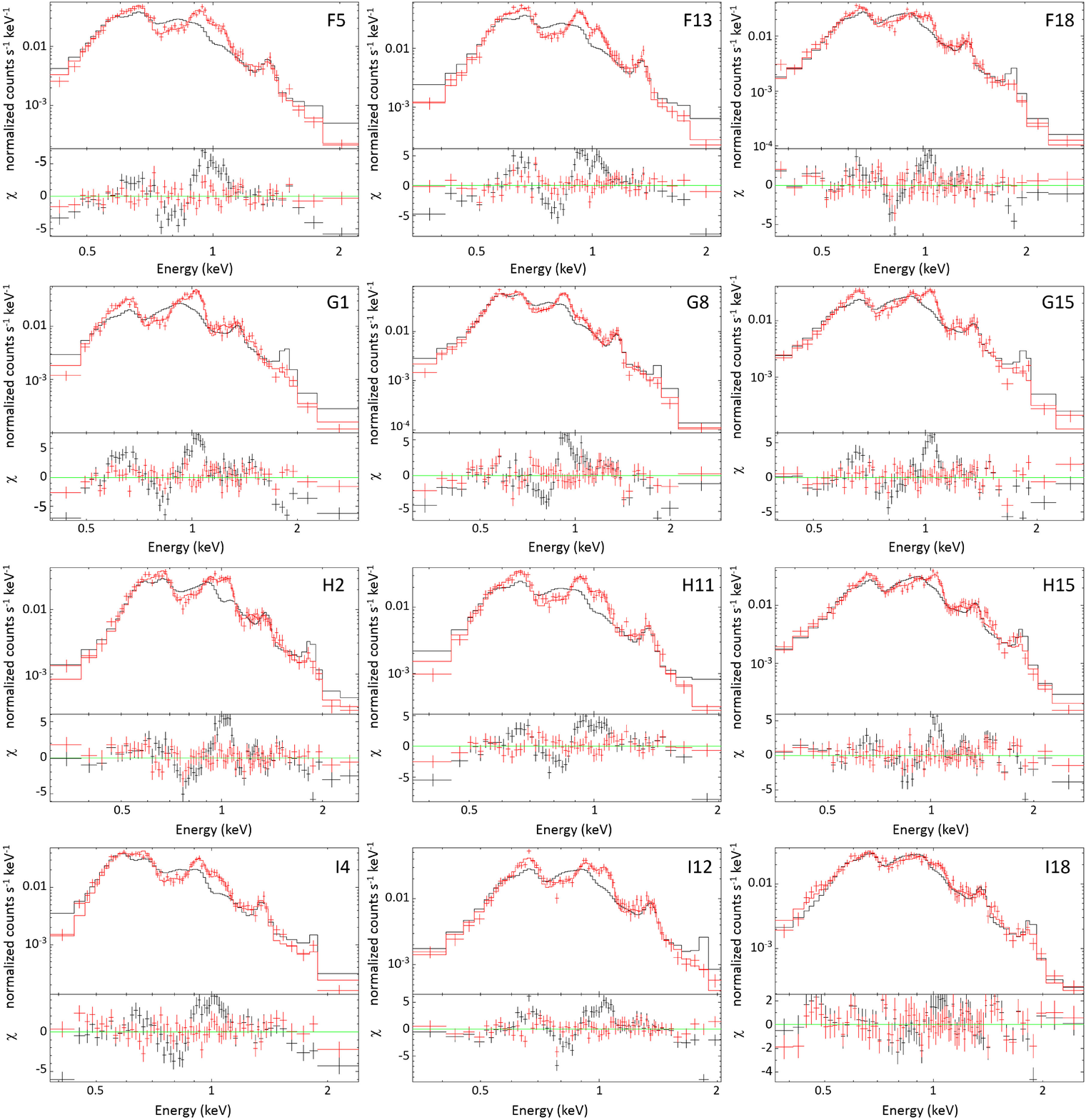}}
\figcaption[]{A small sample of the observed X-ray spectra extracted from regions shown in Fig. 5b. The best-fit one-component shock model with SMC abundances fixed model \citep{RD92} are overlaid in each panel.}
\end{figure*}

\begin{figure*}
\figurenum{B1}
\centerline{\includegraphics[angle=0,width=\textwidth]{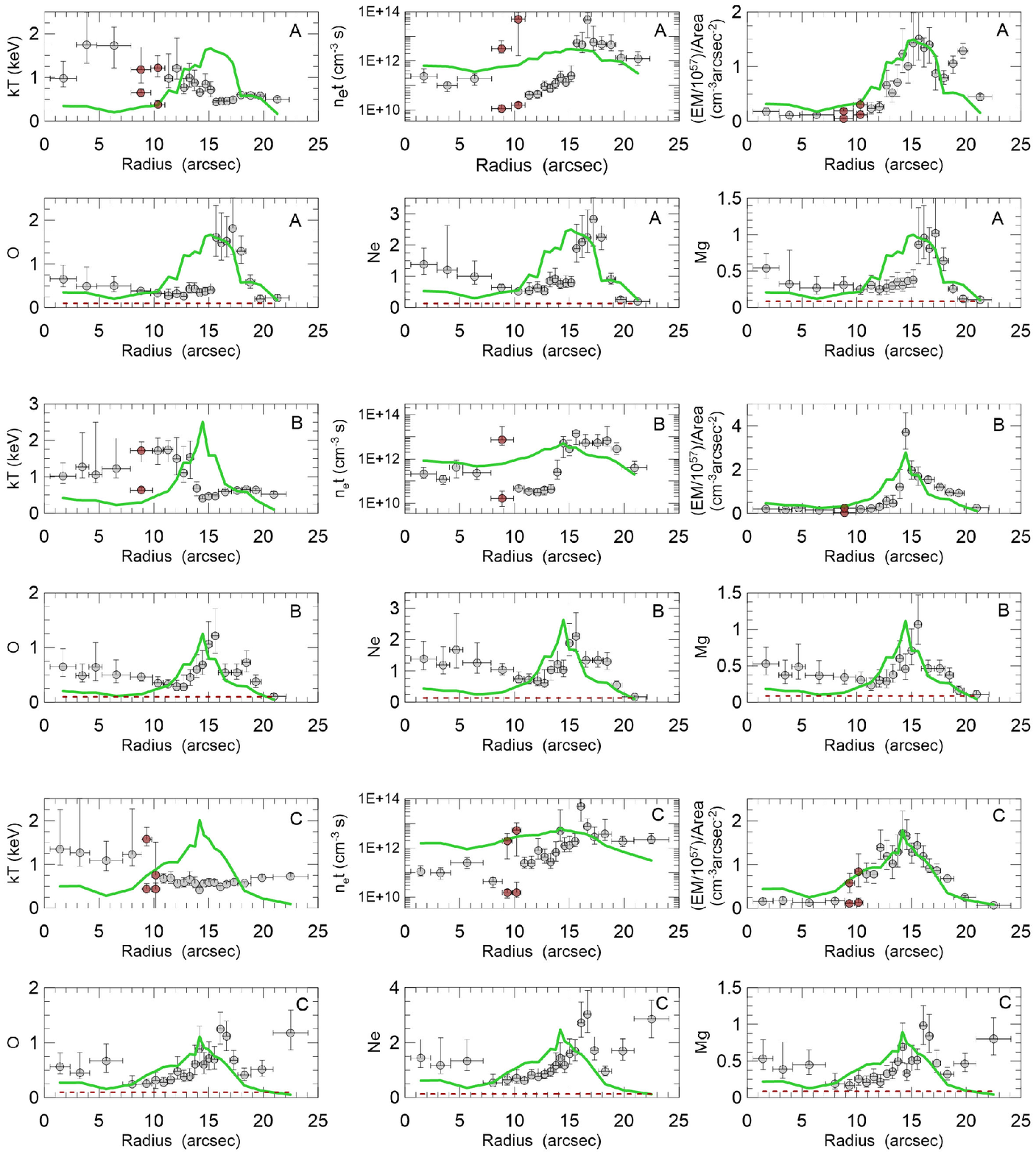}}
\caption[]{{The best-fit spectral parameters with error bars along the radius of E0102. The electron temperature ($\it kT$), ionization timescale ($\it n_{e}t$), emission measure ($\it EM$), O, Ne and Mg abundances in the A (north), B (northwest) and C (west) directions of E0102. Gray and red circles represent results of one-component and two-component shock models, respectively. In each plot the broadband surface brightness profile is overlaid with green line (since the SNR is bright, typical errors for intensity plots are negligible). The red dashed lines in the abundances panels are the CSM abundances for each element. Abundances are with respect to solar \citep{ande89}.}\label{fig:fig8}}
\end{figure*}

\begin{figure*}[]
\figurenum{B2}
\centerline{\includegraphics[angle=0,width=\textwidth]{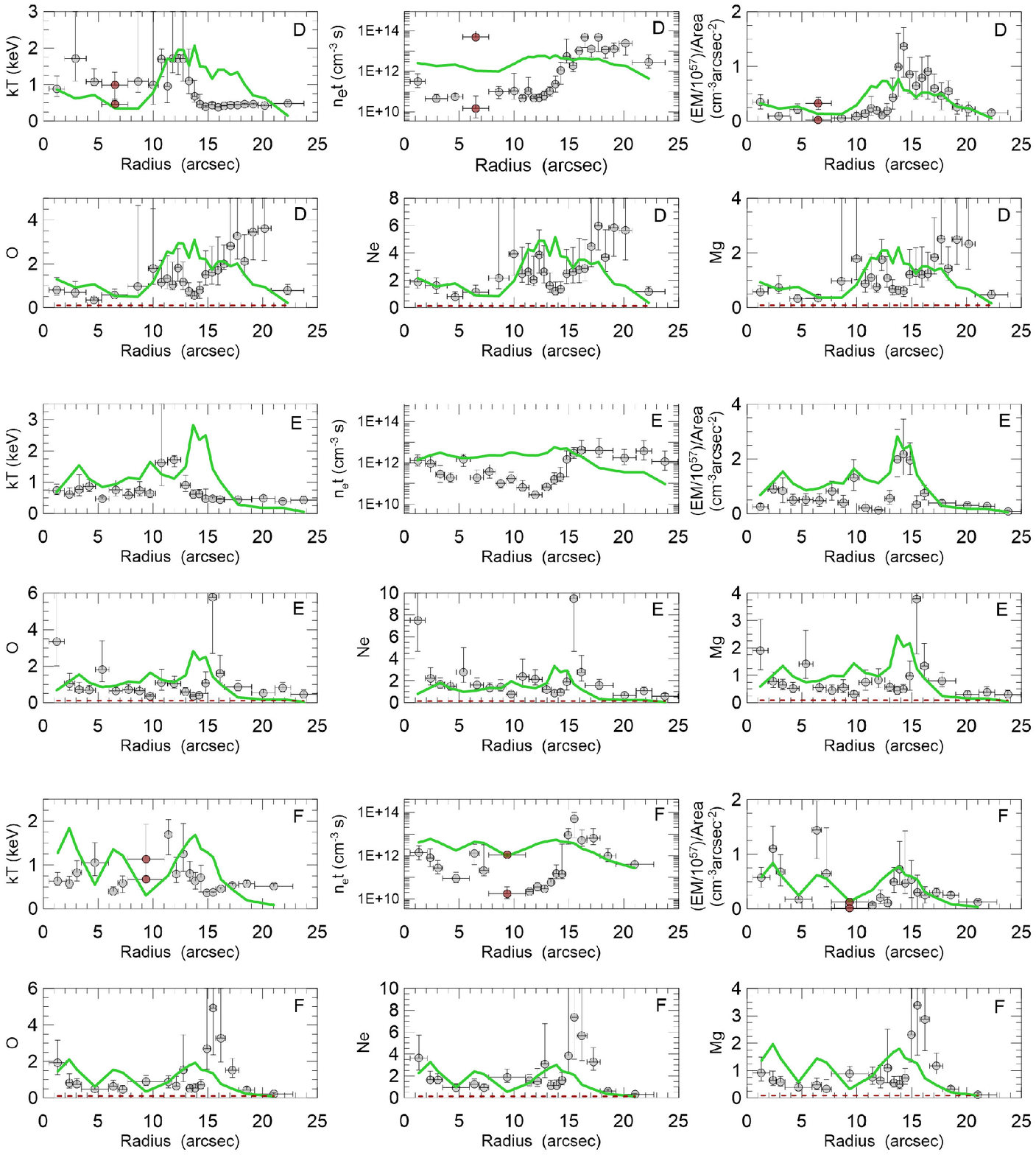}}
\caption[]{{The best-fit spectral parameters with error bars along the radius of E0102. The electron temperature ($\it kT$), ionization timescale ($\it n_{e}t$), emission measure ($\it EM$), O, Ne and Mg abundances in the D (southwest), E (southwestern ridge) and F (south) directions of E0102. Gray and red circles represent results of one-component and two-component shock models, respectively. In each plot the broadband surface brightness profile is overlaid with green line (since the SNR is bright, typical errors for intensity plots are negligible). The red dashed lines in the abundances panels are the CSM abundances for each element. Abundances are with respect to solar \citep{ande89}.}\label{fig:fig8}}
\end{figure*}

\begin{figure*}[]
\figurenum{B3}
\centerline{\includegraphics[angle=0,width=\textwidth]{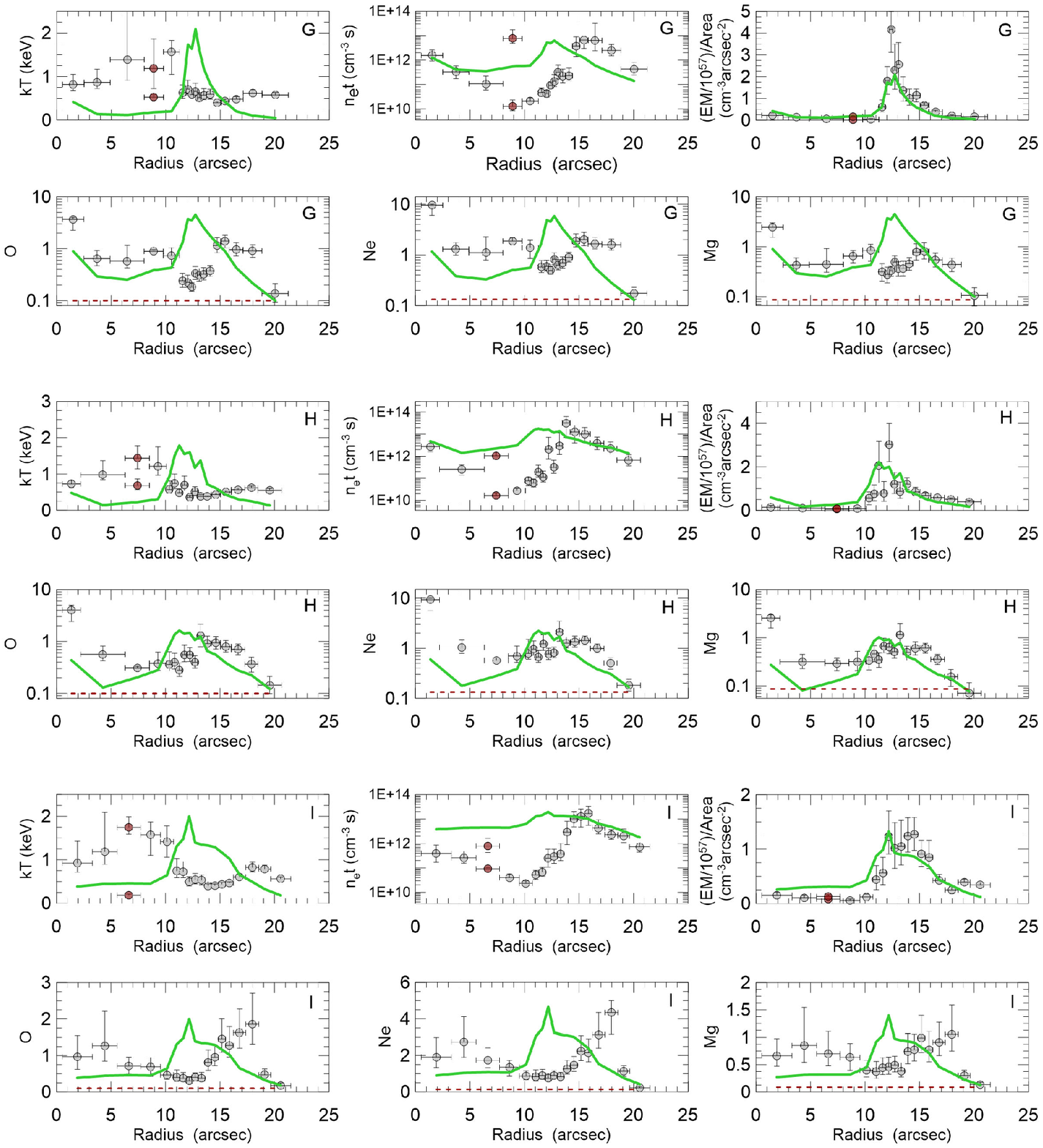}}
\caption[]{{The best-fit spectral parameters with error bars along the radius of E0102. The electron temperature ($\it kT$), ionization timescale ($\it n_{e}t$), emission measure ($\it EM$), O, Ne and Mg abundances in the G (southeast), H (east) and I (northeast) directions of E0102. Gray and red circles represent results of one-component and two-component shock models, respectively. In each plot the broadband surface brightness profile is overlaid with green line (since the SNR is bright, typical errors for intensity plots are negligible). The red dashed lines in the abundances panels are the CSM abundances for each element. Abundances are with respect to solar \citep{ande89}.}\label{fig:fig8}}
\end{figure*}

\newpage

\begin{deluxetable*}{ccccccccccc}
\footnotesize
\tabletypesize{\scriptsize}
\setlength{\tabcolsep}{0.05in}
\tablecaption{Summary of Spectral Model Fits to Radial Regions in the north (A1--A21) direction of E0102}
\label{tbl:tab1}
\tablewidth{0pt}
\tablehead{ \colhead{} & Distance from the & \colhead{} & \colhead{} & \colhead{} & \colhead{} & \colhead{} & \colhead{} & \colhead{} & \colhead{} \\
\colhead{} & Geometric Center & \colhead{$kT$} & \colhead{$n_et$} & \colhead{\it EM} & \colhead{$\chi_{\nu}^2$} & \colhead{O} & \colhead{Ne} & \colhead{Mg} & \colhead{Si} \\
\colhead{Region} & of the SNR (\farcs) & \colhead{(keV)} & \colhead{(10$^{11}$ cm$^{-3}$ s)} & \colhead{($10^{57}$ cm$^{-3}$)} & \colhead{} & \colhead{} & \colhead{} & \colhead{} & \colhead{} & \colhead{}}
\startdata
A 1 & 1.72 & $0.98^{+0.39}_{-0.20}$ & $2.37^{+2.34}_{-1.08}$ & $1.46^{+0.49}_{-0.51}$ & 1.11 & $0.65^{+0.32}_{-0.18}$ & $1.38^{+0.52}_{-0.32}$ & $0.54^{+0.20}_{-0.14}$  & $0.13\tablenotemark{a}$ \\	
A 2 & 3.85 & $1.74^{+3.67}_{-0.42}$ & $1.00^{+0.46}_{-0.43}$ & $0.96^{+0.31}_{-0.44}$ & 1.21 & $0.48^{+0.45}_{-0.11}$ & $1.20^{+1.42}_{-0.26}$ & $0.32^{+0.46}_{-0.10}$  & $0.13\tablenotemark{a}$ \\
A 3 & 6.34 & $1.73^{+0.42}_{-0.51}$ & $1.89^{+1.06}_{-0.86}$ & $1.68^{+0.35}_{-0.32}$ & 1.32 & $0.50^{+0.22}_{-0.13}$ & $1.00^{+0.49}_{-0.26}$ & $0.27^{+0.16}_{-0.09}$ & $0.13\tablenotemark{a}$ \\	
A 4\tablenotemark{b} & 8.80 & $1.18^{+0.51}_{-0.30}$ & $0.11^{+0.04}_{-0.03}$ & $0.46^{+0.08}_{-0.06}$ & 1.43 & $0.38^{+0.02}_{-0.02}$ & $0.64^{+0.09}_{-0.08}$ & $0.31^{+0.11}_{-0.10}$  & $0.13\tablenotemark{a}$ \\
A 4\tablenotemark{b} & 8.80 & $0.65^{+0.51}_{-0.30}$ & $31.58^{+34.13}_{-10.81}$ & $1.64^{+0.31}_{-0.26}$ & 1.43 & $0.38^{+0.02}_{-0.02}$ & $0.64^{+0.09}_{-0.08}$ & $0.31^{+0.11}_{-0.10}$  & $0.13\tablenotemark{a}$ \\
A 5\tablenotemark{b} & 10.33 & $1.22^{+0.28}_{-0.20}$ & $0.16^{+0.04}_{-0.03}$ & $0.95^{+0.10}_{-0.08}$ & 1.14 & $0.33^{+0.02}_{-0.03}$ & $0.52^{+0.07}_{-0.06}$ & $0.25^{+0.08}_{-0.07}$  & $0.13\tablenotemark{a}$ \\	
A 5\tablenotemark{b} & 10.33 & $0.37^{+0.10}_{-0.04}$ & $49.33^{+24.72}_{-14.86}$ & $2.38^{+0.84}_{-0.89}$ & 1.14 & $0.33^{+0.02}_{-0.03}$ & $0.52^{+0.07}_{-0.06}$ & $0.25^{+0.09}_{-0.07}$  & $0.13\tablenotemark{a}$ \\	
A 6 & 11.32 & $0.98^{+0.56}_{-0.21}$ & $0.40^{+0.18}_{-0.13}$ & $0.97^{+0.45}_{-0.41}$ & 1.255 & $0.27^{+0.15}_{-0.07}$ & $0.54^{+0.26}_{-0.14}$ & $0.31^{+0.14}_{-0.10}$  & $0.13\tablenotemark{a}$ \\	
A 7 & 12.08 & $1.21^{+0.69}_{-0.25}$ & $0.44^{+0.16}_{-0.14}$ & $1.22^{+0.44}_{-0.48}$ & 1.364 & $0.32^{+0.14}_{-0.07}$ & $0.63^{+0.20}_{-0.14}$ & $0.25^{+0.11}_{-0.08}$ & $0.13\tablenotemark{a}$ \\
A 8 & 12.73 & $0.78^{+0.21}_{-0.17}$ & $0.92^{+0.56}_{-0.30}$ & $1.63^{+0.69}_{-0.53}$ & 1.272 & $0.26^{+0.09}_{-0.06}$ & $0.53^{+0.18}_{-0.12}$ & $0.27^{+0.11}_{-0.09}$ & $0.13\tablenotemark{a}$ \\
A 9 & 13.23 & $0.98^{+0.33}_{-0.21}$ & $0.76^{+0.41}_{-0.23}$ & $1.30^{+0.56}_{-0.42}$ & 1.484 & $0.43^{+0.14}_{-0.10}$ & $0.86^{+0.28}_{-0.19}$ & $0.30^{+0.12}_{-0.09}$ & $0.13\tablenotemark{a}$ \\	
A 10 & 13.71 & $0.88^{+0.29}_{-0.18}$ & $1.16^{+0.80}_{-0.42}$ & $1.61^{+0.73}_{-0.57}$ & 1.295 & $0.43^{+0.14}_{-0.09}$ & $0.93^{+0.32}_{-0.20}$ & $0.35^{+0.14}_{-0.10}$ & $0.13\tablenotemark{a}$ \\	
A 11 & 14.19 & $0.66^{+0.14}_{-0.11}$ & $2.09^{+1.63}_{-0.76}$ & $2.80^{+1.04}_{-0.78}$ & 1.121 & $0.35^{+0.09}_{-0.07}$ & $0.75^{+0.18}_{-0.14}$ & $0.31^{+0.10}_{-0.08}$ & $0.13\tablenotemark{a}$ \\	
A 12 & 14.68 & $0.84^{+0.23}_{-0.16}$ & $1.36^{+0.93}_{-0.48}$ & $1.89^{+0.73}_{-0.58}$ & 1.114 & $0.36^{+0.10}_{-0.07}$ & $0.80^{+0.23}_{-0.16}$ & $0.37^{+0.18}_{-0.09}$ & $0.13\tablenotemark{a}$ \\	
A 13 & 15.17 & $0.72^{+0.18}_{-0.17}$ & $2.50^{+3.86}_{-1.08}$ & $2.58^{+0.99}_{-0.73}$ & 1.126 & $0.40^{+0.15}_{-0.09}$ & $0.81^{+0.19}_{-0.15}$ & $0.38^{+0.14}_{-0.10}$ & $0.13\tablenotemark{a}$ \\	
A 14 & 15.66 & $0.44^{+0.03}_{-0.04}$ & $53.01^{+15.76}_{-25.76}$ & $2.75^{+0.86}_{-0.71}$ & 1.23 & $1.61^{+0.72}_{-0.44}$ & $1.89^{+0.78}_{-0.43}$ & $0.87^{+0.50}_{-0.26}$ & $0.13\tablenotemark{a}$ \\	
A 15 & 16.15 & $0.46^{+0.05}_{-0.04}$ & $46.30^{+95.14}_{-25.41}$ & $2.57^{+0.34}_{-0.67}$ & 1.25 & $1.48^{+0.68}_{-0.40}$ & $2.10^{+0.85}_{-0.49}$ & $0.95^{+0.45}_{-0.28}$ & $0.13\tablenotemark{a}$ \\	
A 16 & 16.65 & $0.46^{+0.03}_{-0.01}$ & $466.74^{+180.72}_{-380.73}$ & $2.75^{+0.62}_{-0.67}$ & 1.15 & $1.52^{+0.56}_{-0.36}$ & $2.24^{+0.89}_{-0.47}$ & $0.81^{+0.28}_{-0.22}$ & $0.13\tablenotemark{a}$ \\
A 17 & 17.19 & $0.49^{+0.06}_{-0.04}$ & $57.84^{+207.28}_{-30.92}$ & $2.01^{+0.72}_{-0.70}$ & 1.06 & $1.81^{+1.19}_{-0.60}$ & $2.83^{+1.70}_{-0.85}$ & $1.02^{+0.65}_{-0.34}$ & $0.62^{+0.57}_{-0.34}$ \\
A 18 & 17.93 & $0.60^{+0.07}_{-0.05}$ & $47.11^{+50.65}_{-17.35}$ & $4.13^{+0.68}_{-0.64}$ & 1.08 & $1.29^{+0.35}_{-0.28}$ & $2.25^{+0.37}_{-0.40}$ & $0.64^{+0.17}_{-0.18}$ & $0.13\tablenotemark{a}$ \\	
A 19 & 18.80 & $0.59^{+0.02}_{-0.04}$ & $46.25^{+68.40}_{-20.74}$ & $5.31^{+0.66}_{-0.90}$ & 1.09 & $0.59^{+0.17}_{-0.14}$ & $0.91^{+0.19}_{-0.16}$ & $0.26^{+0.08}_{-0.07}$ & $0.13\tablenotemark{a}$ \\	
A 20 & 19.70 & $0.58^{+0.04}_{-0.04}$ & $13.29^{+12.32}_{-6.20}$ & $7.35^{+0.80}_{-0.76}$ & 1.17 & $0.20^{+0.09}_{-0.07}$ & $0.26^{+0.08}_{-0.06}$ & $0.12^{+0.04}_{-0.04}$ & $0.13\tablenotemark{a}$ \\
A 21 & 21.26 & $0.49^{+0.05}_{-0.04}$ & $2.26^{+11.45}_{-5.63}$ & $7.82^{+0.94}_{-1.20}$ & 1.05 & $0.22^{+0.08}_{-0.07}$ & $0.20^{+0.06}_{-0.05}$ & $0.11^{+0.05}_{-0.04}$ & $0.13\tablenotemark{a}$ 
\enddata
\tablecomments{Abundances are with respect to solar \citep{ande89}. The Galactic column $N_{\rm H,Gal}$ is fixed at 0.45 $\times$ 10$^{21}$ cm$^{-2}$ and the SMC column $N_{\rm H,SMC}$ is fixed at the Shell value 0.8 $\times$ 10$^{21}$ cm$^{-2}$. Fe abundance was fixed at the best-fit Shell value. For comparisons the Russell \& Dopita (1992) values for SMC abundances are O = 0.126, Ne = 0.151, Mg = 0.251, Si = 0.302, Fe = 0.149}
\tablenotetext{a}{Si abundance was fixed at the best-fit Shell value}
\tablenotetext{b}{Two shock model parameters for transition regions A4 and A5}
\end{deluxetable*}

\begin{deluxetable*}{ccccccccccc}
\footnotesize
\tabletypesize{\scriptsize}
\setlength{\tabcolsep}{0.05in}
\tablecaption{Summary of Spectral Model Fits to Radial Regions in the northwest (B1--B19) direction of E0102}
\label{tbl:tab1}
\tablewidth{0pt}
\tablehead{ \colhead{} & Distance from the & \colhead{} & \colhead{} & \colhead{} & \colhead{} & \colhead{} & \colhead{} & \colhead{}  \\
\colhead{} & Geometric Center & \colhead{$kT$} & \colhead{$n_et$} & \colhead{\it EM} & \colhead{$\chi_{\nu}^2$} & \colhead{O} & \colhead{Ne} & \colhead{Mg}  \\
\colhead{Region} & of the SNR (\farcs) & \colhead{(keV)} & \colhead{(10$^{11}$ cm$^{-3}$ s)} & \colhead{($10^{57}$ cm$^{-3}$)} & \colhead{} & \colhead{} & \colhead{} & \colhead{} & \colhead{} & \colhead{}}
\startdata
B 1 & 1.70 & $1.01^{+0.36}_{-0.20}$ & $2.09^{+1.97}_{-0.77}$ & $1.10^{+0.42}_{-0.39}$ & 1.12 & $0.65^{+0.32}_{-0.18}$ & $1.38^{+0.56}_{-0.34}$ & $0.53^{+0.23}_{-0.15}$ \\
B 2 & 3.48 & $1.27^{+0.94}_{-0.31}$ & $1.22^{+0.54}_{-0.50}$ & $1.10^{+0.32}_{-0.47}$ & 1.05 & $0.48^{+0.22}_{-0.11}$ & $1.18^{+0.59}_{-0.28}$ & $0.37^{+0.21}_{-0.11}$ \\
B 3 & 4.68 & $1.06^{+1.44}_{-0.23}$ & $4.21^{+4.29}_{-2.84}$ & $1.46^{+0.53}_{-0.95}$ & 1.35 & $0.64^{+0.45}_{-0.25}$ & $1.68^{+1.17}_{-0.59}$ & $0.49^{+0.31}_{-0.17}$ \\
B 4 & 6.57 & $1.22^{+0.84}_{-0.25}$ & $2.32^{+2.01}_{-1.13}$ & $1.49^{+0.51}_{-0.61}$ & 1.23 & $0.50^{+0.27}_{-0.14}$ & $1.26^{+0.64}_{-0.33}$ & $0.37^{+0.20}_{-0.11}$ \\
B 5\tablenotemark{a} & 8.88 & $1.72^{+0.19}_{-0.35}$ & $0.17^{+0.04}_{-0.03}$ & $0.23^{+0.03}_{-0.02}$ & 1.36 & $0.46^{+0.04}_{-0.04}$ & $1.04^{+0.14}_{-0.14}$ & $0.34^{+0.11}_{-0.10}$ \\
B 5\tablenotemark{a} & 8.88 & $0.63^{+0.07}_{-0.06}$ & $72.90^{+21.06}_{-31.06}$ & $2.00^{+0.36}_{-0.31}$ & 1.36 & $0.46^{+0.04}_{-0.04}$ & $1.04^{+0.14}_{-0.14}$ & $0.34^{+0.11}_{-0.10}$ \\
B 6 & 10.38 & $1.72^{+0.35}_{-0.37}$ & $0.47^{+0.17}_{-0.08}$ & $0.92^{+0.27}_{-0.19}$ & 1.24 & $0.36^{+0.11}_{-0.07}$ & $0.75^{+0.22}_{-0.13}$ & $0.30^{+0.11}_{-0.08}$ \\
B 7 & 11.31 & $1.74^{+0.24}_{-0.34}$ & $0.35^{+0.09}_{-0.05}$ & $0.88^{+0.20}_{-0.16}$ & 1.13 & $0.33^{+0.09}_{-0.06}$ & $0.72^{+0.18}_{-0.14}$ & $0.24^{+0.10}_{-0.08}$ \\
B 8 & 12.09 & $1.50^{+0.58}_{-0.45}$ & $0.31^{+0.14}_{-0.08}$ & $0.84^{+0.41}_{-0.25}$ & 1.21 & $0.29^{+0.11}_{-0.08}$ & $0.68^{+0.24}_{-0.16}$ & $0.30^{+0.14}_{-0.09}$ \\
B 9 & 12.72 & $1.10^{+0.73}_{-0.25}$ & $0.39^{+0.16}_{-0.16}$ & $1.04^{+0.45}_{-0.55}$ & 1.14 & $0.28^{+0.21}_{-0.07}$ & $0.61^{+0.43}_{-0.14}$ & $0.29^{+0.16}_{-0.09}$ \\
B 10 & 13.32 & $1.53^{+0.44}_{-0.58}$ & $0.42^{+0.30}_{-0.08}$ & $0.89^{+0.70}_{-0.23}$ & 1.05 & $0.46^{+0.14}_{-0.15}$ & $1.04^{+0.30}_{-0.33}$ & $0.38^{+0.14}_{-0.11}$ \\
B 11 & 13.92 & $0.68^{+0.18}_{-0.20}$ & $2.55^{+6.06}_{-1.32}$ & $1.51^{+0.99}_{-0.64}$ & 1.07 & $0.60^{+0.27}_{-0.14}$ & $1.21^{+0.42}_{-0.27}$ & $0.60^{+0.24}_{-0.16}$ \\
B 12 & 14.46 & $0.40^{+0.07}_{-0.03}$ & $50.60^{+19.13}_{-39.43}$ & $3.77^{+0.89}_{-0.80}$ & 1.10 & $0.68^{+0.26}_{-0.18}$ & $1.04^{+0.31}_{-0.22}$ & $0.46^{+0.20}_{-0.14}$ \\
B 13 & 14.99 & $0.47^{+0.05}_{-0.05}$ & $28.24^{+42.30}_{-14.25}$ & $3.23^{+0.71}_{-0.66}$ & 1.11 & $1.06^{+0.42}_{-0.28}$ & $1.89^{+0.62}_{-0.40}$ & $0.71^{+0.30}_{-0.20}$ \\
B 14 & 15.60 & $0.47^{+0.04}_{-0.02}$ & $140.72^{+36.49}_{-96.89}$ & $2.56^{+0.67}_{-0.61}$ & 1.25 & $1.21^{+0.45}_{-0.32}$ & $2.11^{+0.75}_{-0.49}$ & $1.06^{+0.41}_{-0.27}$ \\
B 15 & 16.50 & $0.59^{+0.05}_{-0.05}$ & $52.54^{+72.26}_{-24.22}$ & $4.87^{+0.46}_{-0.61}$ & 1.20 & $0.54^{+0.16}_{-0.12}$ & $1.34^{+0.29}_{-0.22}$ & $0.46^{+0.11}_{-0.09}$ \\
B 16 & 17.58 & $0.62^{+0.05}_{-0.04}$ & $52.54^{+72.26}_{-24.22}$ & $4.87^{+0.46}_{-0.61}$ & 1.25 & $0.54^{+0.16}_{-0.12}$ & $1.34^{+0.29}_{-0.23}$ & $0.46^{+0.11}_{-0.09}$ \\
B 17 & 18.45 & $0.65^{+0.04}_{-0.05}$ & $67.95^{+219.73}_{-30.16}$ & $4.21^{+0.29}_{-0.52}$ & 1.23 & $0.73^{+0.21}_{-0.17}$ & $1.30^{+0.29}_{-0.25}$ & $0.37^{+0.10}_{-0.10}$ \\
B 18 & 19.32 & $0.63^{+0.05}_{-0.04}$ & $27.85^{+25.84}_{-11.99}$ & $6.24^{+0.63}_{-0.66}$ & 1.13 & $0.36^{+0.12}_{-0.10}$ & $0.54^{+0.14}_{-0.12}$ & $0.16^{+0.06}_{-0.05}$ \\
B 19 & 20.94 & $0.52^{+0.07}_{-0.06}$ & $4.04^{+3.80}_{-1.84}$ & $5.85^{+1.25}_{-1.13}$ & 1.15 & $0.11^{+0.05}_{-0.03}$ & $0.17^{+0.05}_{-0.04}$ & $0.11^{+0.05}_{-0.05}$ 
\enddata
\tablecomments{Abundances are with respect to solar \citep{ande89}. The Galactic column $N_{\rm H,Gal}$ is fixed at 0.45 $\times$ 10$^{21}$ cm$^{-2}$ and the SMC column $N_{\rm H,SMC}$ is fixed at the Shell value 0.8 $\times$ 10$^{21}$ cm$^{-2}$. The Si and Fe abundance were fixed at the best-fit Shell value. For comparisons the Russell \& Dopita (1992) values for SMC abundances are O = 0.126, Ne = 0.151, Mg = 0.251, Si = 0.302, Fe = 0.149}
\tablenotetext{a}{Two shock model parameters for transition region B5}
\end{deluxetable*}

\begin{deluxetable*}{ccccccccccc}
\footnotesize
\tabletypesize{\scriptsize}
\setlength{\tabcolsep}{0.05in}
\tablecaption{Summary of Spectral Model Fits to Radial Regions in the west (C1--C22) direction of E0102}
\label{tbl:tab1}
\tablewidth{0pt}
\tablehead{ \colhead{} & Distance from the & \colhead{} & \colhead{} & \colhead{} & \colhead{} & \colhead{} & \colhead{} & \colhead{} & \colhead{} \\
\colhead{} & Geometric Center & \colhead{$kT$} & \colhead{$n_et$} & \colhead{\it EM} & \colhead{$\chi_{\nu}^2$} & \colhead{O} & \colhead{Ne} & \colhead{Mg} & \colhead{Si} \\
\colhead{Region} & of the SNR (\farcs) & \colhead{(keV)} & \colhead{(10$^{11}$ cm$^{-3}$ s)} & \colhead{($10^{57}$ cm$^{-3}$)} & \colhead{} & \colhead{} & \colhead{} & \colhead{} & \colhead{} & \colhead{}}
\startdata
C 1 & 1.44 & $1.35^{+0.90}_{-0.36}$ & $1.11^{+0.79}_{-0.38}$ & $1.00^{+0.50}_{-0.34}$ & 1.15 & $0.56^{+0.25}_{-0.13}$ & $1.45^{+0.67}_{-0.35}$ & $0.53^{+0.26}_{-0.15}$ & $0.13\tablenotemark{a}$ \\
C 2 & 3.25 & $1.27^{+1.51}_{-0.31}$ & $0.97^{+0.61}_{-0.45}$ & $1.18^{+0.55}_{-0.64}$ & 1.16 & $0.45^{+0.37}_{-0.11}$ & $1.17^{+1.00}_{-0.30}$ & $0.39^{+0.37}_{-0.12}$ & $0.30^{+0.17}_{-0.13}$ \\
C 3 & 5.65 & $1.08^{+0.45}_{-0.22}$ & $2.53^{+1.62}_{-1.07}$ & $1.52^{+0.54}_{-0.60}$ & 1.37 & $0.67^{+0.31}_{-0.21}$ & $1.33^{+0.78}_{-0.34}$ & $0.45^{+0.21}_{-0.13}$ & $0.13\tablenotemark{a}$ \\	
C 4 & 8.02 & $1.23^{+1.04}_{-0.26}$ & $0.44^{+0.18}_{-0.20}$ & $1.24^{+0.47}_{-0.60}$ & 1.12 & $0.24^{+0.16}_{-0.03}$ & $0.54^{+0.31}_{-0.12}$ & $0.19^{+0.14}_{-0.07}$ & $0.13\tablenotemark{a}$ \\	
C 5\tablenotemark{b} & 9.34 & $1.58^{+0.27}_{-0.16}$ & $0.15^{+0.05}_{-0.06}$ & $0.50^{+0.14}_{-0.16}$ & 1.21 & $0.26^{+0.08}_{-0.06}$ & $0.65^{+0.20}_{-0.20}$ & $0.17^{+0.08}_{-0.07}$ & $0.13\tablenotemark{a}$ \\	
C 5\tablenotemark{b} & 9.34 & $0.44^{+0.05}_{-0.12}$ & $19.33^{+19.33}_{-15.75}$ & $2.58^{+1.03}_{-0.98}$ & 1.21 & $0.26^{+0.08}_{-0.06}$ & $0.64^{+0.20}_{-0.20}$ & $0.17^{+0.08}_{-0.07}$ & $0.13\tablenotemark{a}$ \\	
C 6\tablenotemark{b} & 10.18 & $0.76^{+2.35}_{-1.11}$ & $0.15^{+0.24}_{-0.06}$ & $0.47^{+0.38}_{-0.18}$ & 1.46 & $0.32^{+0.15}_{-0.15}$ & $0.71^{+0.31}_{-0.18}$ & $0.25^{+0.12}_{-0.08}$ & $0.13\tablenotemark{a}$ \\
C 6\tablenotemark{b} & 10.18 & $0.43^{+0.14}_{-0.05}$ & $53.04^{+28.35}_{-25.92}$ & $3.02^{+1.48}_{-1.22}$ & 1.46 & $0.32^{+0.15}_{-0.15}$ & $0.71^{+0.31}_{-0.18}$ & $0.25^{+0.12}_{-0.08}$ & $0.13\tablenotemark{a}$ \\
C 7 & 10.90 & $0.68^{+0.12}_{-0.10}$ & $2.35^{+1.85}_{-0.83}$ & $2.50^{+0.82}_{-0.61}$ & 1.12 & $0.28^{+0.09}_{-0.06}$ & $0.62^{+0.16}_{-0.12}$ & $0.21^{+0.09}_{-0.07}$ & $0.13\tablenotemark{a}$ \\
C 8 & 11.55 & $0.68^{+0.15}_{-0.12}$ & $2.44^{+2.09}_{-0.95}$ & $2.46^{+0.80}_{-0.66}$ & 1.12 & $0.32^{+0.10}_{-0.07}$ & $0.82^{+0.20}_{-0.15}$ & $0.29^{+0.10}_{-0.08}$ & $0.13\tablenotemark{a}$ \\
C 9 & 12.16 & $0.57^{+0.08}_{-0.11}$ & $8.07^{+16.06}_{-4.54}$ & $4.18^{+1.18}_{-0.81}$ & 1.34 & $0.48^{+0.27}_{-0.14}$ & $0.76^{+0.30}_{-0.09}$ & $0.21^{+0.10}_{-0.07}$ & $0.13\tablenotemark{a}$ \\	
C 10 & 12.75 & $0.58^{+0.11}_{-0.09}$ & $4.30^{+4.64}_{-1.86}$ & $3.22^{+0.98}_{-0.83}$ & 1.21 & $0.38^{+0.12}_{-0.09}$ & $0.86^{+0.21}_{-0.16}$ & $0.33^{+0.11}_{-0.09}$ & $0.13\tablenotemark{a}$ \\	
C 11 & 13.27 & $0.66^{+0.18}_{-0.13}$ & $2.75^{+3.31}_{-1.25}$ & $2.39^{+0.97}_{-0.79}$ & 1.01 & $0.39^{+0.14}_{-0.09}$ & $0.95^{+0.25}_{-0.18}$ & $0.36^{+0.13}_{-0.10}$ & $0.13\tablenotemark{a}$ \\	
C 12 & 13.75 & $0.57^{+0.14}_{-0.10}$ & $6.66^{+11.52}_{-3.83}$ & $3.02^{+0.92}_{-0.82}$ & 1.22 & $0.62^{+0.34}_{-0.19}$ & $1.18^{+0.45}_{-0.27}$ & $0.50^{+0.23}_{-0.13}$ & $0.13\tablenotemark{a}$ \\	
C 13 & 14.17 & $0.41^{+0.05}_{-0.03}$ & $49.88^{+307.39}_{-30.92}$ & $2.77^{+0.80}_{-0.74}$ & 1.10 & $0.89^{+0.40}_{-0.25}$ & $1.45^{+0.55}_{-0.36}$ & $0.70^{+0.32}_{-0.21}$ & $0.13\tablenotemark{a}$ \\	
C 14 & 14.57 & $0.56^{+0.08}_{-0.07}$ & $12.44^{+13.53}_{-5.51}$ & $3.25^{+0.73}_{-0.61}$ & 1.10 & $0.60^{+0.22}_{-0.16}$ & $1.17^{+0.34}_{-0.26}$ & $0.34^{+0.13}_{-0.10}$ & $0.13\tablenotemark{a}$ \\	
C 15 & 15.02 & $0.59^{+0.09}_{-0.09}$ & $13.12^{+16.72}_{-6.10}$ & $2.79^{+0.44}_{-0.54}$ & 1.11 & $0.72^{+0.30}_{-0.21}$ & $1.61^{+0.55}_{-0.39}$ & $0.51^{+0.20}_{-0.14}$ & $0.13\tablenotemark{a}$ \\
C 16 & 15.73 & $0.58^{+0.06}_{-0.06}$ & $18.64^{+20.20}_{-7.29}$ & $3.58^{+0.66}_{-0.57}$ & 1.12 & $0.70^{+0.23}_{-0.18}$ & $1.68^{+0.42}_{-0.35}$ & $0.51^{+0.15}_{-0.13}$ & $0.13\tablenotemark{a}$ \\	
C 17 & 16.07 & $0.48^{+0.03}_{-0.02}$ & $498.56^{+177.12}_{-377.72}$ & $3.03^{+0.61}_{-0.62}$ & 1.34 & $1.25^{+0.31}_{-0.27}$ & $2.71^{+0.77}_{-0.53}$ & $0.98^{+0.27}_{-0.25}$ & $0.46^{+0.20}_{-0.24}$ \\
C 18 & 16.64 & $0.55^{+0.06}_{-0.05}$ & $76.75^{+10.14}_{-40.84}$ & $2.75^{+0.67}_{-0.29}$ & 1.13 & $1.12^{+0.28}_{-0.26}$ & $3.03^{+0.86}_{-0.67}$ & $0.84^{+0.29}_{-0.22}$ & $0.33^{+0.29}_{-0.21}$ \\
C 19 & 17.30 & $0.60^{+0.06}_{-0.07}$ & $28.64^{+41.14}_{-14.29}$ & $3.43^{+0.61}_{-0.55}$ & 1.20 & $0.69^{+0.17}_{-0.18}$ & $1.71^{+0.48}_{-0.34}$ & $0.47^{+0.17}_{-0.13}$ & $0.47^{+0.28}_{-0.19}$ \\
C 20 & 18.26 & $0.56^{+0.04}_{-0.06}$ & $37.32^{+108.25}_{-17.10}$ & $5.46^{+0.84}_{-0.65}$ & 1.11 & $0.42^{+0.13}_{-0.10}$ & $0.96^{+0.19}_{-0.16}$ & $0.32^{+0.10}_{-0.08}$ & $0.25^{+0.16}_{-0.11}$ \\
C 21 & 19.88 & $0.69^{+0.07}_{-0.06}$ & $18.79^{+11.21}_{-7.28}$ & $3.85^{+0.58}_{-0.28}$ & 1.30 & $0.52^{+0.16}_{-0.13}$ & $1.71^{+0.44}_{-0.34}$ & $0.46^{+0.14}_{-0.11}$ & $0.13^{+0.12}_{-0.09}$ \\
C 22 & 22.49 & $0.72^{+0.08}_{-0.07}$ & $22.26^{+17.15}_{-7.46}$ & $2.49^{+0.54}_{-0.52}$ & 1.14 & $1.18^{+0.42}_{-0.30}$ & $2.86^{+0.68}_{-0.69}$ & $0.80^{+0.29}_{-0.21}$ & $0.30^{+0.20}_{-0.09}$ 
\enddata
\tablecomments{Abundances are with respect to solar \citep{ande89}. The Galactic column $N_{\rm H,Gal}$ is fixed at 0.45 $\times$ 10$^{21}$ cm$^{-2}$ and the SMC column $N_{\rm H,SMC}$ is fixed at the Shell value 0.8 $\times$ 10$^{21}$ cm$^{-2}$. Fe abundance was fixed at the best-fit Shell value. For comparisons the Russell \& Dopita (1992) values for SMC abundances are O = 0.126, Ne = 0.151, Mg = 0.251, Si = 0.302, Fe = 0.149}
\tablenotetext{a}{Si abundance was fixed at the best-fit Shell value}
\tablenotetext{b}{Two shock model parameters for transition regions C5 and C6}
\end{deluxetable*}

\begin{deluxetable*}{ccccccccccc}
\footnotesize
\tabletypesize{\scriptsize}
\setlength{\tabcolsep}{0.05in}
\tablecaption{Summary of Spectral Model Fits to Radial Regions in the southwest (D1--D24) direction of E0102}
\label{tbl:tab1}
\tablewidth{0pt}
\tablehead{ \colhead{} & Distance from the & \colhead{} & \colhead{} & \colhead{} & \colhead{} & \colhead{} & \colhead{} & \colhead{}  \\
\colhead{} & Geometric Center & \colhead{$kT$} & \colhead{$n_et$} & \colhead{\it EM} & \colhead{$\chi_{\nu}^2$} & \colhead{O} & \colhead{Ne} & \colhead{Mg} \\
\colhead{Region} & of the SNR (\farcs) & \colhead{(keV)} & \colhead{(10$^{11}$ cm$^{-3}$ s)} & \colhead{($10^{57}$ cm$^{-3}$)} & \colhead{} & \colhead{} & \colhead{} & \colhead{} & \colhead{} & \colhead{}}
\startdata
D 1 & 1.22 & $0.88^{+0.35}_{-0.20}$ & $3.14^{+4.31}_{-1.56}$ & $1.44^{+0.54}_{-0.52}$ & 1.11 & $0.81^{+0.57}_{-0.24}$ & $1.88^{+0.83}_{-0.47}$ & $0.57^{+0.23}_{-0.16}$ \\
D 2 & 2.91 & $1.71^{+2.17}_{-0.62}$ & $0.45^{+0.26}_{-0.15}$ & $0.52^{+0.38}_{-0.24}$ & 1.32 & $0.70^{+0.49}_{-0.22}$ & $1.61^{+0.57}_{-0.56}$ & $0.73^{+0.44}_{-0.21}$ \\
D 3 & 4.62 & $1.08^{+0.35}_{-0.11}$ & $0.55^{+0.25}_{-0.17}$ & $1.03^{+0.49}_{-0.37}$ & 1.41 & $0.35^{+0.10}_{-0.10}$ & $0.81^{+0.34}_{-0.21}$ & $0.33^{+0.15}_{-0.10}$ \\
D 4 \tablenotemark{a} & 6.51 & $0.98^{+0.36}_{-0.38}$ & $0.15^{+0.46}_{-0.10}$ & $0.21^{+0.30}_{-0.10}$ & 1.51 & $0.58^{+0.26}_{-0.15}$ & $1.13^{+0.52}_{-0.28}$ & $0.34^{+0.14}_{-0.11}$ \\
D 4 \tablenotemark{a} & 6.51 & $0.46^{+0.09}_{-0.04}$ & $50.00^{+5.46}_{-38.22}$ & $3.01^{+1.00}_{-1.00}$ & 1.51 & $0.58^{+0.26}_{-0.15}$ & $1.13^{+0.52}_{-0.28}$ & $0.34^{+0.14}_{-0.11}$ \\
D 5 & 8.62 & $1.09^{+3.60}_{-0.31}$ & $0.96^{+0.82}_{-0.52}$ & $0.56^{+0.51}_{-0.56}$ & 1.32 & $0.99^{+3.68}_{-0.35}$ & $2.16^{+8.20}_{-0.79}$ & $0.97^{+3.37}_{-0.36}$ \\
D 6 & 10.00 & $0.98^{+4.03}_{-0.53}$ & $1.08^{+6.81}_{-0.67}$ & $0.36^{+0.35}_{-0.28}$ & 1.12 & $1.78^{+2.73}_{-0.72}$ & $3.93^{+6.30}_{-1.56}$ & $1.79^{+2.91}_{-0.76}$ \\
D 7 & 10.76 & $1.70^{+0.28}_{-0.48}$ & $0.48^{+0.19}_{-0.15}$ & $0.38^{+0.18}_{-0.16}$ & 1.43 & $1.17^{+0.99}_{-0.29}$ & $2.47^{+1.94}_{-0.62}$ & $0.88^{+0.52}_{-0.32}$ \\
D 8 & 11.29 & $0.95^{+0.74}_{-0.45}$ & $1.05^{+4.02}_{-0.50}$ & $0.48^{+0.43}_{-0.26}$ & 1.23 & $1.34^{+1.03}_{-0.48}$ & $2.63^{+2.07}_{-0.96}$ & $1.10^{+0.85}_{-0.40}$ \\
D 9 & 11.79 & $1.71^{+2.45}_{-0.39}$ & $0.47^{+0.13}_{-0.06}$ & $0.44^{+0.14}_{-0.11}$ & 1.12 & $1.05^{+1.03}_{-0.23}$ & $2.04^{+1.73}_{-0.44}$ & $0.76^{+0.42}_{-0.20}$ \\
D 10 & 12.28 & $1.71^{+0.45}_{-0.40}$ & $0.49^{+0.08}_{-0.07}$ & $0.26^{+0.08}_{-0.08}$ & 1.22 & $1.46^{+0.70}_{-0.39}$ & $3.20^{+1.50}_{-0.83}$ & $1.36^{+0.72}_{-0.41}$ \\
D 11 & 12.74 & $1.71^{+2.41}_{-0.62}$ & $0.64^{+0.42}_{-0.26}$ & $0.36^{+0.25}_{-0.19}$ & 1.13 & $1.18^{+0.91}_{-0.35}$ & $2.60^{+1.94}_{-0.79}$ & $1.08^{+0.79}_{-0.34}$ \\
D 12 & 13.26 & $1.10^{+0.86}_{-0.35}$ & $1.04^{+0.66}_{-0.43}$ & $1.04^{+0.86}_{-0.51}$ & 1.35 & $0.75^{+0.40}_{-0.21}$ & $1.62^{+0.93}_{-0.50}$ & $0.68^{+0.35}_{-0.21}$ \\
D 13 & 13.77 & $0.68^{+0.43}_{-0.16}$ & $2.40^{+3.31}_{-1.46}$ & $1.64^{+1.02}_{-0.90}$ & 1.37 & $0.55^{+0.22}_{-0.14}$ & $1.22^{+0.65}_{-0.29}$ & $0.65^{+0.14}_{-0.17}$ \\
D 14 & 14.26 & $0.47^{+0.26}_{-0.07}$ & $11.22^{+17.26}_{-8.93}$ & $3.22^{+0.80}_{-1.33}$ & 1.13 & $0.82^{+0.35}_{-0.39}$ & $1.38^{+0.50}_{-0.36}$ & $0.61^{+0.19}_{-0.22}$ \\
D 15 & 14.80 & $0.39^{+0.03}_{-0.02}$ & $54.84^{+338.87}_{-29.58}$ & $2.04^{+0.74}_{-0.81}$ & 1.35 & $1.50^{+1.09}_{-0.47}$ & $2.48^{+1.76}_{-0.73}$ & $1.21^{+0.91}_{-0.41}$ \\
D 16 & 15.37 & $0.44^{+0.06}_{-0.04}$ & $19.32^{+24.04}_{-9.68}$ & $1.90^{+0.78}_{-0.69}$ & 1.22 & $1.61^{+1.19}_{-0.54}$ & $2.62^{+1.78}_{-0.81}$ & $1.27^{+0.92}_{-0.43}$ \\
D 17 & 15.93 & $0.38^{+0.02}_{-0.02}$ & $105.24^{+17.21}_{-67.21}$ & $1.94^{+0.95}_{-0.84}$ & 1.33 & $1.72^{+1.50}_{-0.68}$ & $2.78^{+2.30}_{-0.95}$ & $1.20^{+1.01}_{-0.45}$ \\
D 18 & 16.45 & $0.41^{+0.01}_{-0.01}$ & $500.00^{+151.15}_{-351.12}$ & $2.22^{+0.69}_{-1.22}$ & 1.52 & $2.00^{+0.86}_{-0.24}$ & $2.87^{+2.08}_{-0.90}$ & $1.24^{+0.77}_{-0.26}$ \\
D 19 & 17.05 & $0.44^{+0.02}_{-0.03}$ & $124.77^{+43.92}_{-734.00}$ & $1.72^{+0.75}_{-0.74}$ & 1.45 & $2.82^{+2.42}_{-0.93}$ & $4.48^{+3.66}_{-1.47}$ & $1.84^{+1.44}_{-0.61}$ \\
D 20 & 17.68 & $0.44^{+0.02}_{-0.01}$ & $500.00^{+93.24}_{-403.23}$ & $1.23^{+0.64}_{-0.68}$ & 1.54 & $3.27^{+4.41}_{-1.05}$ & $5.97^{+7.69}_{-1.75}$ & $2.51^{+3.37}_{-0.91}$ \\
D 21 & 18.32 & $0.47^{+0.02}_{-0.02}$ & $115.13^{+50.99}_{-70.99}$ & $1.94^{+0.66}_{-0.64}$ & 1.24 & $2.11^{+1.22}_{-0.62}$ & $3.67^{+2.02}_{-1.02}$ & $1.43^{+0.79}_{-0.43}$ \\
D 22 & 19.12 & $0.46^{+0.03}_{-0.03}$ & $125.59^{+45.93}_{-75.98}$ & $1.28^{+0.66}_{-0.63}$ & 1.14 & $3.45^{+2.19}_{-1.28}$ & $5.84^{+5.72}_{-2.08}$ & $2.50^{+2.64}_{-0.91}$ \\
D 23 & 20.17 & $0.42^{+0.03}_{-0.01}$ & $240.67^{+118.24}_{-178.74}$ & $1.24^{+0.71}_{-0.71}$ & 1.10 & $3.61^{+5.25}_{-1.46}$ & $5.64^{+7.82}_{-2.17}$ & $2.33^{+3.57}_{-0.92}$ \\
D 24 & 22.27 & $0.48^{+0.06}_{-0.04}$ & $27.73^{+36.00}_{-13.52}$ & $3.81^{+0.92}_{-0.86}$ & 1.06 & $0.79^{+0.27}_{-0.19}$ & $1.17^{+0.36}_{-0.25}$ & $0.47^{+0.16}_{-0.13}$ 
\enddata
\tablecomments{Abundances are with respect to solar \citep{ande89}. The Galactic column $N_{\rm H,Gal}$ is fixed at 0.45 $\times$ 10$^{21}$ cm$^{-2}$ and the SMC column $N_{\rm H,SMC}$ is fixed at the Shell value 0.8 $\times$ 10$^{21}$ cm$^{-2}$. The Si and Fe abundance were fixed at the best-fit Shell value. For comparisons the Russell \& Dopita (1992) values for SMC abundances are O = 0.126, Ne = 0.151, Mg = 0.251, Si = 0.302, Fe = 0.149}
\tablenotetext{a}{Two shock model parameters for transition region D4}
\end{deluxetable*}

\begin{deluxetable*}{ccccccccccc}
\footnotesize
\tabletypesize{\scriptsize}
\setlength{\tabcolsep}{0.05in}
\tablecaption{Summary of Spectral Model Fits to Radial Regions in the bright ridge at the southwest (E1--E21) direction of E0102}
\label{tbl:tab1}
\tablewidth{0pt}
\tablehead{ \colhead{} & Distance from the & \colhead{} & \colhead{} & \colhead{} & \colhead{} & \colhead{} & \colhead{} & \colhead{}  \\
\colhead{} & Geometric Center & \colhead{$kT$} & \colhead{$n_et$} & \colhead{\it EM} & \colhead{$\chi_{\nu}^2$} & \colhead{O} & \colhead{Ne} & \colhead{Mg} \\
\colhead{Region}  & of the SNR (\farcs) & \colhead{(keV)} & \colhead{(10$^{11}$ cm$^{-3}$ s)} & \colhead{($10^{57}$ cm$^{-3}$)} & \colhead{} & \colhead{} & \colhead{} & \colhead{} & \colhead{} & \colhead{}}
\startdata
E 1 & 1.22 & $0.73^{+0.21}_{-0.13}$ & $12.83^{+11.18}_{-5.79}$ & $0.94^{+0.45}_{-0.44}$ & 1.12 & $3.35^{+3.41}_{-1.31}$ & $7.50^{+7.46}_{-2.84}$ & $1.91^{+1.13}_{-0.71}$ \\
E 2 & 2.40 & $0.62^{+0.20}_{-0.10}$ & $9.14^{+10.29}_{-5.67}$ & $1.93^{+0.47}_{-0.27}$ & 1.26 & $1.05^{+0.55}_{-0.36}$ & $2.22^{+0.95}_{-0.61}$ & $0.78^{+0.34}_{-0.22}$ \\
E 3 & 3.27 & $0.77^{+0.47}_{-0.20}$ & $2.78^{+4.58}_{-1.63}$ & $1.43^{+0.80}_{-0.75}$ & 1.11 & $0.72^{+0.40}_{-0.18}$ & $1.67^{+0.56}_{-0.38}$ & $0.65^{+0.20}_{-0.17}$ \\
E 4 & 4.20 & $0.86^{+0.24}_{-0.16}$ & $1.89^{+1.53}_{-0.72}$ & $1.12^{+0.49}_{-0.42}$ & 1.24 & $0.68^{+0.26}_{-0.16}$ & $1.48^{+0.59}_{-0.33}$ & $0.52^{+0.23}_{-0.15}$ \\
E 5 & 5.38 & $0.46^{+0.06}_{-0.05}$ & $15.16^{+10.82}_{-8.38}$ & $1.53^{+0.68}_{-0.62}$ & 1.10 & $1.81^{+1.58}_{-0.66}$ & $2.78^{+2.24}_{-0.94}$ & $1.42^{+1.22}_{-0.53}$ \\
E 6 & 6.62 & $0.75^{+0.31}_{-0.20}$ & $1.88^{+2.67}_{-0.88}$ & $1.34^{+0.71}_{-0.57}$ & 1.33 & $0.64^{+0.22}_{-0.15}$ & $1.61^{+0.62}_{-0.38}$ & $0.56^{+0.23}_{-0.16}$ \\
E 7 & 7.76 & $0.59^{+0.18}_{-0.12}$ & $3.70^{+4.58}_{-1.91}$ & $1.86^{+0.84}_{-0.73}$ & 1.24 & $0.71^{+0.30}_{-0.18}$ & $1.33^{+0.47}_{-0.30}$ & $0.45^{+0.20}_{-0.14}$ \\
E 8 & 8.79 & $0.74^{+0.28}_{-0.18}$ & $1.01^{+0.91}_{-0.40}$ & $0.93^{+0.62}_{-0.40}$ & 1.54 & $0.64^{+0.31}_{-0.18}$ & $1.42^{+0.67}_{-0.38}$ & $0.55^{+0.29}_{-0.18}$ \\
E 9 & 9.75 & $0.64^{+0.17}_{-0.13}$ & $1.72^{+1.84}_{-0.70}$ & $2.07^{+1.04}_{-0.72}$ & 1.14 & $0.35^{+0.11}_{-0.08}$ & $0.76^{+0.24}_{-0.16}$ & $0.31^{+0.13}_{-0.10}$ \\
E 10 & 10.79 & $1.62^{+2.11}_{-0.74}$ & $0.64^{+0.75}_{-0.22}$ & $0.44^{+0.53}_{-0.21}$ & 1.36 & $1.08^{+0.76}_{-0.39}$ & $2.36^{+1.58}_{-0.90}$ & $0.74^{+0.45}_{-0.28}$ \\
E 11 & 11.95 & $1.71^{+0.14}_{-0.22}$ & $0.28^{+0.04}_{-0.03}$ & $0.31^{+0.02}_{-0.01}$ & 1.36 & $1.04^{+0.41}_{-0.24}$ & $2.12^{+0.85}_{-0.50}$ & $0.82^{+0.41}_{-0.28}$ \\
E 12 & 12.98 & $0.91^{+0.31}_{-0.21}$ & $0.67^{+0.40}_{-0.20}$ & $1.07^{+0.59}_{-0.40}$ & 1.36 & $0.59^{+0.25}_{-0.15}$ & $1.20^{+0.50}_{-0.31}$ & $0.57^{+0.25}_{-0.16}$ \\
E 13 & 13.67 & $0.61^{+0.14}_{-0.13}$ & $1.53^{+1.73}_{-0.59}$ & $1.78^{+0.99}_{-0.57}$ & 1.24 & $0.34^{+0.10}_{-0.08}$ & $0.87^{+0.24}_{-0.20}$ & $0.44^{+0.16}_{-0.13}$ \\
E 14 & 14.23 & $0.62^{+0.15}_{-0.08}$ & $2.10^{+3.91}_{-0.85}$ & $2.37^{+1.40}_{-0.83}$ & 1.51 & $0.39^{+0.12}_{-0.08}$ & $0.94^{+0.24}_{-0.19}$ & $0.50^{+0.17}_{-0.12}$ \\
E 15 & 14.82 & $0.46^{+0.31}_{-0.06}$ & $14.78^{+24.11}_{-12.66}$ & $2.15^{+0.69}_{-0.58}$ & 1.21 & $1.06^{+0.64}_{-0.56}$ & $1.89^{+0.96}_{-0.74}$ & $0.97^{+0.55}_{-0.51}$ \\
E 16 & 15.44 & $0.47^{+0.06}_{-0.04}$ & $28.80^{+25.90}_{-12.24}$ & $0.64^{+0.62}_{-0.64}$ & 1.31 & $5.76^{+415.48}_{-3.06}$ & $9.50^{+365.63}_{-4.84}$ & $3.79^{+397.06}_{-2.01}$ \\
E 17 & 16.14 & $0.44^{+0.04}_{-0.03}$ & $42.43^{+82.33}_{-21.85}$ & $2.02^{+0.76}_{-0.69}$ & 1.22 & $1.60^{+1.00}_{-0.51}$ & $2.76^{+1.54}_{-0.80}$ & $1.34^{+0.81}_{-0.42}$ \\
E 18 & 17.76 & $0.44^{+0.04}_{-0.03}$ & $40.04^{+109.89}_{-21.64}$ & $3.08^{+0.85}_{-0.76}$ & 1.44 & $0.86^{+0.35}_{-0.23}$ & $1.56^{+0.56}_{-0.35}$ & $0.79^{+0.31}_{-0.21}$ \\
E 19 & 20.07 & $0.48^{+0.09}_{-0.06}$ & $17.46^{+26.95}_{-9.41}$ & $3.78^{+0.90}_{-0.79}$ & 1.21 & $0.53^{+0.20}_{-0.16}$ & $0.67^{+0.20}_{-0.15}$ & $0.30^{+0.13}_{-0.10}$ \\
E 20 & 21.84 & $0.38^{+0.03}_{-0.03}$ & $38.22^{+80.85}_{-24.55}$ & $3.71^{+0.98}_{-1.40}$ & 1.33 & $0.82^{+0.30}_{-0.22}$ & $1.06^{+0.35}_{-0.22}$ & $0.38^{+0.20}_{-0.14}$ \\
E 21 & 23.76 & $0.43^{+0.11}_{-0.06}$ & $11.84^{+26.48}_{-7.52}$ & $2.54^{+1.27}_{-0.95}$ & 1.16 & $0.46^{+0.26}_{-0.16}$ & $0.55^{+0.26}_{-0.16}$ & $0.29^{+0.17}_{-0.13}$ 
\enddata
\tablecomments{Abundances are with respect to solar \citep{ande89}. The Galactic column $N_{\rm H,Gal}$ is fixed at 0.45 $\times$ 10$^{21}$ cm$^{-2}$ and the SMC column $N_{\rm H,SMC}$ is fixed at the Shell value 0.8 $\times$ 10$^{21}$ cm$^{-2}$. The Si and Fe abundance were fixed at the best-fit Shell value. For comparisons the Russell \& Dopita (1992) values for SMC abundances are O = 0.126, Ne = 0.151, Mg = 0.251, Si = 0.302, Fe = 0.149}
\end{deluxetable*}

\begin{deluxetable*}{ccccccccccc}
\footnotesize
\tabletypesize{\scriptsize}
\setlength{\tabcolsep}{0.05in}
\tablecaption{Summary of Spectral Model Fits to Radial Regions in the south (F1--F19) direction of E0102}
\label{tbl:tab1}
\tablewidth{0pt}
\tablehead{ \colhead{} & Distance from the & \colhead{} & \colhead{} & \colhead{} & \colhead{} & \colhead{} & \colhead{} & \colhead{}  \\
\colhead{}  & Geometric Center & \colhead{$kT$} & \colhead{$n_et$} & \colhead{\it EM} & \colhead{$\chi_{\nu}^2$} & \colhead{O} & \colhead{Ne} & \colhead{Mg}\\
\colhead{Region} & of the SNR (\farcs) & \colhead{(keV)} & \colhead{(10$^{11}$ cm$^{-3}$ s)} & \colhead{($10^{57}$ cm$^{-3}$)} & \colhead{} & \colhead{} & \colhead{} & \colhead{} & \colhead{} & \colhead{}}
\startdata
F 1 & 1.31 & $0.63^{+0.20}_{-0.10}$ & $14.19^{+13.75}_{-7.72}$ & $1.50^{+0.50}_{-0.47}$ & 1.05 & $1.93^{+1.22}_{-0.65}$ & $3.63^{+2.09}_{-1.09}$ & $0.92^{+0.56}_{-0.30}$ \\
F 2 & 2.37 & $0.57^{+0.18}_{-0.11}$ & $7.83^{+12.6}_{-4.77}$ & $1.90^{+0.71}_{-0.57}$ & 1.07 & $0.83^{+0.50}_{-0.29}$ & $1.67^{+0.78}_{-0.45}$ & $0.65^{+0.35}_{-0.20}$ \\
F 3 & 3.06 & $0.83^{+0.27}_{-0.20}$ & $2.67^{+3.47}_{-1.19}$ & $1.65^{+0.78}_{-0.62}$ & 1.02 & $0.75^{+0.31}_{-0.18}$ & $1.66^{+0.54}_{-0.36}$ & $0.58^{+0.21}_{-0.15}$ \\
F 4 & 4.71 & $1.05^{+0.46}_{-0.30}$ & $0.90^{+0.76}_{-0.33}$ & $0.90^{+0.52}_{-0.32}$ & 1.24 & $0.48^{+0.18}_{-0.12}$ & $0.93^{+0.36}_{-0.24}$ & $0.38^{+0.17}_{-0.12}$ \\
F 5 & 6.38 & $0.39^{+0.11}_{-0.04}$ & $12.98^{+21.24}_{-9.12}$ & $3.58^{+1.33}_{-1.29}$ & 1.53 & $0.65^{+0.32}_{-0.20}$ & $1.23^{+0.52}_{-0.31}$ & $0.48^{+0.25}_{-0.16}$ \\
F 6 & 7.25 & $0.59^{+0.16}_{-0.22}$ & $2.13^{+30.18}_{-0.96}$ & $1.68^{+2.16}_{-0.62}$ & 1.26 & $0.46^{+0.22}_{-0.12}$ & $0.93^{+0.34}_{-0.21}$ & $0.32^{+0.34}_{-0.12}$ \\
F 7 \tablenotemark{a} & 9.37 & $1.13^{+0.80}_{-0.47}$ & $0.18^{+0.20}_{-0.07}$ & $0.16^{+0.07}_{-0.03}$ & 1.23 & $0.89^{+0.10}_{-0.11}$ & $1.87^{+0.26}_{-0.33}$ & $0.88^{+0.22}_{-0.19}$ \\
F 7 \tablenotemark{a} & 9.37 & $0.68^{+0.07}_{-0.07}$ & $10.82^{+1.19}_{-0.99}$ & $1.37^{+0.30}_{-0.23}$ & 1.23 & $0.89^{+0.10}_{-0.11}$ & $1.87^{+0.26}_{-0.33}$ & $0.88^{+0.22}_{-0.19}$ \\
F 8 & 11.41 & $1.70^{+0.34}_{-0.46}$ & $0.22^{+0.06}_{-0.03}$ & $0.30^{+0.16}_{-0.30}$ & 1.21 & $0.88^{+0.34}_{-0.33}$ & $1.56^{+0.58}_{-0.47}$ & $0.75^{+0.37}_{-0.25}$ \\
F 9 & 12.13 & $0.79^{+0.39}_{-0.19}$ & $0.36^{+0.21}_{-0.12}$ & $0.59^{+0.41}_{-0.30}$ & 1.21 & $0.65^{+0.54}_{-0.22}$ & $1.52^{+1.17}_{-0.48}$ & $0.63^{+0.56}_{-0.27}$ \\
F 10 & 12.79 & $1.26^{+0.69}_{-0.40}$ & $0.30^{+0.15}_{-0.08}$ & $0.25^{+0.25}_{-0.15}$ & 1.01 & $1.53^{+1.93}_{-0.64}$ & $3.09^{+3.70}_{-1.24}$ & $1.10^{+1.41}_{-0.54}$ \\
F 11 & 13.36 & $0.80^{+0.28}_{-0.17}$ & $0.60^{+0.31}_{-0.19}$ & $1.06^{+0.56}_{-0.41}$ & 1.09 & $0.52^{+0.23}_{-0.14}$ & $1.13^{+0.47}_{-0.28}$ & $0.55^{+0.24}_{-0.16}$ \\
F 12 & 13.87 & $0.63^{+0.20}_{-0.18}$ & $1.51^{+2.26}_{-0.64}$ & $1.38^{+0.98}_{-0.52}$ & 1.33 & $0.54^{+0.24}_{-0.14}$ & $1.14^{+0.49}_{-0.29}$ & $0.51^{+0.29}_{-0.18}$ \\
F 13 & 14.38 & $0.71^{+0.28}_{-0.33}$ & $1.46^{+33.23}_{-0.58}$ & $1.08^{+2.22}_{-0.43}$ & 1.13 & $0.70^{+0.31}_{-0.19}$ & $1.62^{+0.70}_{-0.43}$ & $0.73^{+0.35}_{-0.22}$ \\
F 14 & 14.95 & $0.36^{+0.02}_{-0.02}$ & $89.59^{+24.28}_{-54.78}$ & $1.40^{+0.92}_{-0.86}$ & 1.13 & $2.70^{+3.37}_{-1.13}$ & $3.85^{+6.66}_{-1.57}$ & $2.31^{+4.08}_{-1.00}$ \\
F 15 & 15.49 & $0.37^{+0.01}_{-0.01}$ & $500.00^{+132.17}_{-432.47}$ & $0.83^{+0.86}_{-0.83}$ & 1.12 & $4.93^{+652.15}_{-2.56}$ & $7.34^{+597.88}_{-3.83}$ & $3.39^{+308.96}_{-1.82}$ \\
F 16 & 16.18 & $0.45^{+0.04}_{-0.03}$ & $51.79^{+102.66}_{-24.64}$ & $1.22^{+0.75}_{-0.72}$ & 1.21 & $3.28^{+5.03}_{-1.31}$ & $5.67^{+7.59}_{-2.24}$ & $2.87^{+4.09}_{-1.15}$ \\
F 17 & 17.23 & $0.54^{+0.05}_{-0.04}$ & $63.58^{+115.69}_{-30.80}$ & $2.28^{+0.62}_{-0.59}$ & 1.12 & $1.52^{+0.65}_{-0.40}$ & $3.27^{+1.32}_{-0.75}$ & $1.17^{+0.46}_{-0.29}$ \\
F 18 & 18.55 & $0.57^{+0.07}_{-0.08}$ & $9.88^{+11.15}_{-4.64}$ & $4.08^{+0.91}_{-0.68}$ & 1.10 & $0.44^{+0.18}_{-0.14}$ & $0.62^{+0.19}_{-0.15}$ & $0.32^{+0.10}_{-0.09}$ \\
F 19 & 21.01 & $0.51^{+0.06}_{-0.06}$ & $3.94^{+3.35}_{-1.51}$ & $5.26^{+1.36}_{-1.08}$ & 1.08 & $0.22^{+0.07}_{-0.05}$ & $0.31^{+0.07}_{-0.06}$ & $0.12^{+0.06}_{-0.05}$ 
\enddata
\tablecomments{Abundances are with respect to solar \citep{ande89}. The Galactic column $N_{\rm H,Gal}$ is fixed at 0.45 $\times$ 10$^{21}$ cm$^{-2}$ and the SMC column $N_{\rm H,SMC}$ is fixed at the Shell value 0.8 $\times$ 10$^{21}$ cm$^{-2}$. The Si and Fe abundance were fixed at the best-fit Shell value. For comparisons the Russell \& Dopita (1992) values for SMC abundances are O = 0.126, Ne = 0.151, Mg = 0.251, Si = 0.302, Fe = 0.149}
\tablenotetext{a}{Two shock model parameters for spectrum of region F7}
\end{deluxetable*}

\begin{deluxetable*}{ccccccccccc}
\footnotesize
\tabletypesize{\scriptsize}
\setlength{\tabcolsep}{0.05in}
\tablecaption{Summary of Spectral Model Fits to Radial Regions in the southeast (G1--G17) direction of E0102}
\label{tbl:tab1}
\tablewidth{0pt}
\tablehead{ \colhead{} & Distance from the & \colhead{} & \colhead{} & \colhead{} & \colhead{} & \colhead{} & \colhead{} & \colhead{} & \colhead{} \\
\colhead{} & Geometric Center & \colhead{$kT$} & \colhead{$n_et$} & \colhead{\it EM} & \colhead{$\chi_{\nu}^2$} & \colhead{O} & \colhead{Ne} & \colhead{Mg} & \colhead{Si} \\
\colhead{Region} & of the SNR (\farcs) & \colhead{(keV)} & \colhead{(10$^{11}$ cm$^{-3}$ s)} & \colhead{($10^{57}$ cm$^{-3}$)} & \colhead{} & \colhead{} & \colhead{} & \colhead{} & \colhead{} & \colhead{}}
\startdata
G 1 &  1.51 & $0.82^{+0.23}_{-0.14}$ & $15.39^{+10.73}_{-6.12}$ & $0.90^{+0.44}_{-0.90}$ & 1.13 & $3.68^{+5.83}_{-1.42}$ & $9.61^{+14.84}_{-3.56}$ & $2.45^{+3.96}_{-0.91}$ & $0.13\tablenotemark{a}$ \\	
G 2 & 3.70 &  $0.87^{+0.30}_{-0.14}$ & $3.26^{+2.50}_{-1.53}$ & $1.75^{+0.54}_{-0.59}$ & 1.11 & $0.64^{+0.28}_{-0.17}$ & $1.30^{+0.44}_{-0.31}$ & $0.43^{+0.17}_{-0.12}$ & $0.13\tablenotemark{a}$ \\
G 3 &  6.46 & $1.38^{+2.52}_{-0.47}$ & $1.06^{+1.15}_{-0.47}$ & $1.07^{+0.74}_{-0.50}$ & 1.25 & $0.58^{+0.58}_{-0.15}$ & $1.11^{+1.16}_{-0.32}$ & $0.44^{+0.47}_{-0.14}$ & $0.13\tablenotemark{a}$ \\
G 4\tablenotemark{b} &  8.90 & $1.18^{+0.68}_{-0.46}$ & $0.13^{+0.11}_{-0.04}$ & $0.18^{+0.06}_{-0.03}$ & 1.25 & $0.89^{+0.07}_{-0.04}$ & $1.87^{+0.30}_{-0.25}$ & $0.65^{+0.19}_{-0.17}$ & $0.13\tablenotemark{a}$ \\	
G 4\tablenotemark{b} &  8.90 & $0.52^{+0.68}_{-0.05}$ & $76.54^{+96.86}_{-27.41}$ & $1.56^{+0.27}_{-0.24}$ & 1.25 & $0.89^{+0.07}_{-0.04}$ & $1.87^{+0.30}_{-0.25}$ & $0.65^{+0.19}_{-0.17}$ & $0.13\tablenotemark{a}$ \\	
G 5 &  10.52 & $1.56^{+0.26}_{-0.51}$ & $0.21^{+0.07}_{-0.03}$ & $0.33^{+0.28}_{-0.10}$ & 1.34 & $0.74^{+0.30}_{-0.30}$ & $1.39^{+0.56}_{-0.54}$ & $0.84^{+0.28}_{-0.33}$ & $0.13\tablenotemark{a}$ \\
G 6 &  11.56 & $0.63^{+0.18}_{-0.12}$ & $0.47^{+0.26}_{-0.14}$ & $1.78^{+0.94}_{-0.66}$ & 1.11 & $0.24^{+0.10}_{-0.06}$ & $0.58^{+0.20}_{-0.13}$ & $0.32^{+0.14}_{-0.11}$ & $0.13\tablenotemark{a}$ \\	
G 7 &  12.04 & $0.71^{+0.20}_{-0.11}$ & $0.42^{+0.16}_{-0.12}$ & $1.77^{+0.64}_{-0.61}$ & 1.24 & $0.22^{+0.08}_{-0.04}$ & $0.60^{+0.19}_{-0.11}$ & $0.28^{+0.11}_{-0.09}$ & $0.13\tablenotemark{a}$ \\	
G 8 &  12.39 & $0.58^{+0.09}_{-0.06}$ & $0.92^{+0.33}_{-0.24}$ & $4.20^{+1.07}_{-1.05}$ & 1.12 & $0.18^{+0.04}_{-0.03}$ & $0.49^{+0.10}_{-0.07}$ & $0.33^{+0.08}_{-0.07}$ & $0.13\tablenotemark{a}$ \\
G 9 &  12.72 & $0.66^{+0.19}_{-0.13}$ & $1.37^{+1.20}_{-0.55}$ & $1.86^{+0.98}_{-0.71}$ & 1.12 & $0.33^{+0.12}_{-0.08}$ & $0.82^{+0.29}_{-0.18}$ & $0.49^{+0.17}_{-0.12}$ & $0.13\tablenotemark{a}$ \\	
G 10 &  13.07 & $0.51^{+0.12}_{-0.08}$ & $3.15^{+3.06}_{-1.48}$ & $2.84^{+1.11}_{-1.01}$ & 1.12 & $0.31^{+0.10}_{-0.07}$ & $0.63^{+0.18}_{-0.12}$ & $0.36^{+0.13}_{-0.11}$ & $0.13\tablenotemark{a}$ \\	
G 11 &  13.50 & $0.57^{+0.15}_{-0.10}$ & $2.15^{+2.15}_{-0.96}$ & $2.17^{+1.00}_{-0.79}$ & 1.22 & $0.33^{+0.10}_{-0.07}$ & $0.70^{+0.21}_{-0.15}$ & $0.37^{+0.14}_{-0.11}$ & $0.13\tablenotemark{a}$ \\
G 12 &  14.07 & $0.59^{+0.11}_{-0.11}$ & $2.28^{+2.47}_{-0.82}$ & $2.23^{+1.00}_{-0.68}$ & 1.23 & $0.37^{+0.10}_{-0.08}$ & $0.89^{+0.23}_{-0.17}$ & $0.46^{+0.15}_{-0.12}$ & $0.13\tablenotemark{a}$ \\	
G 13 &  14.72 & $0.40^{+0.03}_{-0.02}$ & $37.85^{+51.85}_{-23.78}$ & $3.40^{+0.87}_{-0.91}$ & 1.36 & $1.15^{+0.47}_{-0.31}$ & $1.85^{+0.75}_{-0.41}$ & $0.78^{+0.36}_{-0.24}$ & $0.13\tablenotemark{a}$ \\	
G 14 &  15.44 & $0.44^{+0.05}_{-0.03}$ & $65.56^{+15.61}_{-35.61}$ & $2.88^{+0.48}_{-0.76}$ & 1.11 & $1.41^{+0.41}_{-0.37}$ & $2.01^{+0.79}_{-0.49}$ & $0.81^{+0.39}_{-0.24}$ & $0.13\tablenotemark{a}$ \\	
G 15 &  16.46 & $0.48^{+0.06}_{-0.03}$ & $63.89^{+250.56}_{-39.84}$ & $3.34^{+0.55}_{-0.80}$ & 1.36 & $0.96^{+0.35}_{-0.24}$ & $1.65^{+0.57}_{-0.33}$ & $0.55^{+0.20}_{-0.16}$ & $0.13\tablenotemark{a}$ \\	
G 16 &  17.97 & $0.62^{+0.08}_{-0.06}$ & $24.98^{+18.50}_{-9.88}$ & $3.42^{+0.62}_{-0.30}$ & 1.24 & $0.90^{+0.29}_{-0.22}$ & $1.59^{+0.46}_{-0.34}$ & $0.44^{+0.15}_{-0.12}$ & $0.30^{+0.18}_{-0.15}$ \\
G 17 &  20.04 & $0.57^{+0.06}_{-0.05}$ & $4.32^{+3.83}_{-1.87}$ & $5.74^{+0.90}_{-0.85}$ & 1.31 & $0.14^{+0.08}_{-0.04}$ & $0.18^{+0.06}_{-0.04}$ & $0.10^{+0.04}_{-0.04}$ & $0.13\tablenotemark{a}$ 	
\enddata
\tablecomments{Abundances are with respect to solar \citep{ande89}. The Galactic column $N_{\rm H,Gal}$ is fixed at 0.45 $\times$ 10$^{21}$ cm$^{-2}$ and the SMC column $N_{\rm H,SMC}$ is fixed at the Shell value 0.8 $\times$ 10$^{21}$ cm$^{-2}$. Fe abundance was fixed at the best-fit Shell value. For comparisons the Russell \& Dopita (1992) values for SMC abundances are O = 0.126, Ne = 0.151, Mg = 0.251, Si = 0.302, Fe = 0.149}
\tablenotetext{a}{Si abundance was fixed at the best-fit Shell value}
\tablenotetext{b}{Two shock model parameters for transition region G4}
\end{deluxetable*}

\begin{deluxetable*}{ccccccccccc}
\footnotesize
\tabletypesize{\scriptsize}
\setlength{\tabcolsep}{0.05in}
\tablecaption{Summary of Spectral Model Fits to Radial Regions in the east (H1--H17) direction of E0102}
\label{tbl:tab1}
\tablewidth{0pt}
\tablehead{ \colhead{} & Distance from the & \colhead{} & \colhead{} & \colhead{} & \colhead{} & \colhead{} & \colhead{} & \colhead{} & \colhead{} \\
\colhead{} & Geometric Center & \colhead{$kT$} & \colhead{$n_et$} & \colhead{\it EM} & \colhead{$\chi_{\nu}^2$} & \colhead{O} & \colhead{Ne} & \colhead{Mg} & \colhead{Si} \\
\colhead{Region} & of the SNR (\farcs) & \colhead{(keV)} & \colhead{(10$^{11}$ cm$^{-3}$ s)} & \colhead{($10^{57}$ cm$^{-3}$)} & \colhead{} & \colhead{} & \colhead{} & \colhead{} & \colhead{} & \colhead{}}
\startdata
H 1 & 1.36 & $0.73^{+0.10}_{-0.10}$ & $26.91^{+21.34}_{-10.94}$ & $0.80^{+0.45}_{-0.80}$ & 1.41 & $4.03^{+5.11}_{-1.61}$ & $9.31^{+12.35}_{-3.70}$ & $2.55^{+3.22}_{-0.97}$ & $0.13\tablenotemark{a}$ \\	
H 2 & 4.25 & $0.98^{+0.39}_{-0.13}$ & $2.59^{+1.46}_{-1.19}$ & $1.93^{+0.45}_{-0.68}$ & 1.31 & $0.57^{+0.25}_{-0.14}$ & $1.04^{+0.42}_{-0.25}$ & $0.32^{+0.14}_{-0.09}$ & $0.13\tablenotemark{a}$ \\	
H 3\tablenotemark{b}& 7.42 & $1.43^{+0.35}_{-0.29}$ & $0.17^{+0.05}_{-0.04}$ & $0.68^{+0.08}_{-0.06}$ & 1.24 & $0.31^{+0.02}_{-0.02}$ & $0.57^{+0.07}_{-0.06}$ & $0.29^{+0.08}_{-0.08}$ & $0.13\tablenotemark{a}$ \\	
H 3\tablenotemark{b} & 7.42 & $0.68^{+0.18}_{-0.11}$ & $10.42^{+5.94}_{-2.85}$ & $1.01^{+0.32}_{-0.29}$ & 1.24 & $0.31^{+0.02}_{-0.02}$ & $0.57^{+0.07}_{-0.06}$ & $0.29^{+0.08}_{-0.08}$ & $0.13\tablenotemark{a}$ \\
H 4 & 9.31  & $1.21^{+0.54}_{-0.24}$ & $0.28^{+0.07}_{-0.07}$ & $0.76^{+0.32}_{-0.33}$ & 1.32 & $0.38^{+0.24}_{-0.10}$ & $0.70^{+0.41}_{-0.17}$ & $0.32^{+0.20}_{-0.11}$ & $0.13\tablenotemark{a}$ \\	
H 5 & 10.36 & $0.57^{+0.24}_{-0.09}$ & $0.77^{+0.37}_{-0.32}$ & $1.54^{+0.75}_{-0.85}$ & 1.41 & $0.36^{+0.28}_{-0.09}$ & $0.79^{+0.72}_{-0.18}$ & $0.34^{+0.22}_{-0.12}$ & $0.13\tablenotemark{a}$ \\	
H 6 & 10.84 & $0.74^{+0.26}_{-0.15}$ & $0.61^{+0.32}_{-0.21}$ & $1.30^{+0.70}_{-0.54}$ & 1.21 & $0.39^{+0.20}_{-0.10}$ & $0.96^{+0.44}_{-0.23}$ & $0.46^{+0.22}_{-0.14}$ & $0.13\tablenotemark{a}$ \\	
H 7 & 11.27 & $0.48^{+0.11}_{-0.08}$ & $1.95^{+1.98}_{-0.79}$ & $3.16^{+1.68}_{-1.22}$ & 1.13 & $0.28^{+0.11}_{-0.07}$ & $0.67^{+0.23}_{-0.14}$ & $0.35^{+0.14}_{-0.10}$ & $0.13\tablenotemark{a}$ \\	
H 8 & 11.73 & $0.69^{+0.25}_{-0.15}$ & $1.14^{+0.98}_{-0.46}$ & $1.28^{+0.88}_{-0.59}$ & 1.16 & $0.56^{+0.29}_{-0.15}$ & $1.23^{+0.64}_{-0.32}$ & $0.67^{+0.31}_{-0.18}$ & $0.13\tablenotemark{a}$ \\	
H 9 & 12.21 & $0.36^{+0.06}_{-0.02}$ & $20.51^{+51.62}_{-13.02}$ & $5.12^{+1.63}_{-1.34}$ & 1.18 & $0.56^{+0.20}_{-0.14}$ & $0.76^{+0.24}_{-0.16}$ & $0.65^{+0.25}_{-0.18}$ & $0.13\tablenotemark{a}$ \\	
H 10 & 12.69 & $0.52^{+0.13}_{-0.09}$ & $3.13^{+3.46}_{-1.45}$ & $2.83^{+1.24}_{-1.01}$ & 1.13 & $0.40^{+0.13}_{-0.09}$ & $0.81^{+0.24}_{-0.16}$ & $0.51^{+0.17}_{-0.13}$ & $0.13\tablenotemark{a}$ \\	
H 11 & 13.20 & $0.38^{+0.05}_{-0.03}$ & $29.94^{+85.02}_{-18.15}$ & $1.76^{+0.86}_{-0.64}$ & 1.24 & $1.29^{+0.87}_{-0.50}$ & $2.11^{+1.37}_{-0.72}$ & $1.15^{+0.82}_{-0.45}$ & $0.13\tablenotemark{a}$ \\	
H 12 & 13.83 & $0.38^{+0.01}_{-0.01}$ & $317.54^{+147.45}_{-247.45}$ & $4.34^{+1.02}_{-0.88}$ & 1.14 & $0.90^{+0.36}_{-0.22}$ & $1.26^{+0.41}_{-0.25}$ & $0.50^{+0.17}_{-0.16}$ & $0.13\tablenotemark{a}$ \\
H 13 & 14.61 & $0.43^{+0.03}_{-0.02}$ & $125.33^{+45.63}_{-85.63}$ & $3.64^{+0.93}_{-0.71}$ & 1.23 & $0.94^{+0.31}_{-0.22}$ & $1.35^{+0.40}_{-0.29}$ & $0.60^{+0.22}_{-0.16}$ & $0.13\tablenotemark{a}$ \\	
H 14 & 15.53 & $0.50^{+0.03}_{-0.03}$ & $101.47^{+29.75}_{-59.76}$ & $4.24^{+0.80}_{-0.69}$ & 1.34 & $0.79^{+0.25}_{-0.17}$ & $1.42^{+0.33}_{-0.26}$ & $0.61^{+0.16}_{-0.13}$ & $0.13\tablenotemark{a}$ \\	
H 15 & 16.66 & $0.57^{+0.04}_{-0.02}$ & $39.48^{+34.33}_{-19.50}$ & $5.05^{+0.71}_{-0.71}$ & 1.22 & $0.70^{+0.19}_{-0.15}$ & $0.99^{+0.23}_{-0.17}$ & $0.36^{+0.10}_{-0.08}$ & $0.13\tablenotemark{a}$ \\
H 16 & 17.88 & $0.62^{+0.05}_{-0.05}$ & $22.32^{+21.46}_{-9.78}$ & $5.41^{+0.63}_{-0.63}$ & 1.13 & $0.36^{+0.13}_{-0.11}$ & $0.50^{+0.14}_{-0.12}$ & $0.15^{+0.06}_{-0.06}$ & $0.13^{+0.10}_{-0.05}$ \\
H 17 & 19.56 & $0.55^{+0.05}_{-0.06}$ & $6.60^{+5.09}_{-2.91}$ & $7.86^{+1.37}_{-0.89}$ & 1.09 & $0.14^{+0.07}_{-0.04}$ & $0.18^{+0.06}_{-0.04}$ & $0.07^{+0.04}_{-0.03}$ & $0.28^{+0.13}_{-0.10}$ 
\enddata
\tablecomments{Abundances are with respect to solar \citep{ande89}. The Galactic column $N_{\rm H,Gal}$ is fixed at 0.45 $\times$ 10$^{21}$ cm$^{-2}$ and the SMC column $N_{\rm H,SMC}$ is fixed at the Shell value 0.8 $\times$ 10$^{21}$ cm$^{-2}$. Fe abundance was fixed at the best-fit Shell value. For comparisons the Russell \& Dopita (1992) values for SMC abundances are O = 0.126, Ne = 0.151, Mg = 0.251, Si = 0.302, Fe = 0.149}
\tablenotetext{a}{Si abundance was fixed at the best-fit Shell value}
\tablenotetext{b}{Two shock model parameters for transition region H3}
\end{deluxetable*}

\begin{deluxetable*}{ccccccccccc}
\footnotesize
\tabletypesize{\scriptsize}
\setlength{\tabcolsep}{0.05in}
\tablecaption{Summary of Spectral Model Fits to Radial Regions in the northeast (I1--I18) direction of E0102}
\label{tbl:tab1}
\tablewidth{0pt}
\tablehead{ \colhead{} & Distance from the & \colhead{} & \colhead{} & \colhead{} & \colhead{} & \colhead{} & \colhead{} & \colhead{} & \colhead{} \\
\colhead{} & Geometric Center & \colhead{$kT$} & \colhead{$n_et$} & \colhead{\it EM} & \colhead{$\chi_{\nu}^2$} & \colhead{O} & \colhead{Ne} & \colhead{Mg} & \colhead{Si} \\
\colhead{Region} & of the SNR (\farcs) & \colhead{(keV)} & \colhead{(10$^{11}$ cm$^{-3}$ s)} & \colhead{($10^{57}$ cm$^{-3}$)} & \colhead{} & \colhead{} & \colhead{} & \colhead{} & \colhead{} & \colhead{}}
\startdata
I 1  & 1.91  & $0.92^{+0.50}_{-0.21}$ & $3.98^{+4.68}_{-2.28}$ & $1.44^{+0.50}_{-0.56}$ & 1.22 & $0.97^{+0.58}_{-0.35}$ & $1.92^{+1.06}_{-0.58}$ & $0.67^{+0.31}_{-0.19}$  & $0.13\tablenotemark{a}$ \\	
I 2  & 4.42  & $1.18^{+0.90}_{-0.29}$ & $2.57^{+2.27}_{-1.01}$ & $0.92^{+0.40}_{-0.47}$ & 1.24 & $1.27^{+0.94}_{-0.42}$ & $2.74^{+1.38}_{-0.85}$ & $0.85^{+0.69}_{-0.31}$  & $0.13\tablenotemark{a}$ \\		
I 3\tablenotemark{b}& 6.63 & $1.74^{+0.24}_{-0.37}$ & $0.92^{+0.10}_{-0.09}$ & $0.63^{+0.05}_{-0.04}$ & 1.31 & $0.71^{+0.25}_{-0.22}$ & $1.74^{+0.68}_{-0.41}$ & $0.70^{+0.40}_{-0.22}$  & $0.13\tablenotemark{a}$ \\
I 3\tablenotemark{b}& 6.63 & $0.19^{+0.02}_{-0.02}$ & $7.85^{+8.35}_{-3.56}$ & $1.04^{+0.27}_{-0.21}$ & 1.31 & $0.71^{+0.25}_{-0.22}$ & $1.74^{+0.68}_{-0.41}$ & $0.70^{+0.40}_{-0.22}$  & $0.13\tablenotemark{a}$ \\		
I 4  & 8.62  & $1.58^{+0.29}_{-0.47}$ & $0.40^{+0.17}_{-0.08}$ & $0.50^{+0.19}_{-0.13}$ & 1.21 & $0.69^{+0.25}_{-0.27}$ & $1.36^{+0.37}_{-0.50}$ & $0.63^{+0.25}_{-0.24}$  & $0.13\tablenotemark{a}$ \\		
I 5  & 10.13 & $1.41^{+0.36}_{-0.36}$ & $0.24^{+0.06}_{-0.05}$ & $0.69^{+0.29}_{-0.21}$ & 1.01 & $0.47^{+0.19}_{-0.12}$ & $0.87^{+0.28}_{-0.20}$ & $0.40^{+0.18}_{-0.12}$  & $0.13\tablenotemark{a}$ \\		
I 6  & 11.03 & $0.74^{+0.29}_{-0.16}$ & $0.52^{+0.27}_{-0.17}$ & $1.23^{+0.70}_{-0.52}$ & 1.03 & $0.40^{+0.20}_{-0.11}$ & $0.83^{+0.37}_{-0.21}$ & $0.37^{+0.19}_{-0.13}$  & $0.13\tablenotemark{a}$ \\		
I 7  & 11.63 & $0.73^{+0.23}_{-0.14}$ & $0.67^{+0.37}_{-0.22}$ & $1.48^{+0.76}_{-0.57}$ & 1.09 & $0.38^{+0.16}_{-0.09}$ & $0.90^{+0.34}_{-0.20}$ & $0.45^{+0.18}_{-0.13}$  & $0.13\tablenotemark{a}$ \\		
I 8  & 12.17 & $0.49^{+0.12}_{-0.09}$ & $2.50^{+3.34}_{-1.43}$ & $2.41^{+1.36}_{-0.95}$ & 1.09 & $0.30^{+0.11}_{-0.08}$ & $0.75^{+0.26}_{-0.16}$ & $0.46^{+0.18}_{-0.13}$  & $0.13\tablenotemark{a}$ \\		
I 9  & 12.71 & $0.54^{+0.14}_{-0.09}$ & $2.88^{+2.93}_{-1.37}$ & $2.73^{+1.24}_{-1.00}$ & 1.03 & $0.40^{+0.12}_{-0.09}$ & $0.91^{+0.26}_{-0.17}$ & $0.49^{+0.16}_{-0.12}$  & $0.13\tablenotemark{a}$ \\		
I 10 & 13.31 & $0.53^{+0.12}_{-0.10}$ & $3.76^{+6.32}_{-1.78}$ & $2.86^{+1.30}_{-0.93}$ & 1.12 & $0.38^{+0.16}_{-0.09}$ & $0.83^{+0.24}_{-0.16}$ & $0.38^{+0.15}_{-0.11}$  & $0.13\tablenotemark{a}$ \\		
I 11 & 13.91 & $0.40^{+0.05}_{-0.03}$ & $29.37^{+51.86}_{-20.26}$ & $3.44^{+0.93}_{-0.83}$ & 1.29 & $0.82^{+0.34}_{-0.22}$ & $1.27^{+0.47}_{-0.28}$ & $0.74^{+0.33}_{-0.22}$  & $0.13\tablenotemark{a}$ \\		
I 12 & 14.52 & $0.41^{+0.04}_{-0.02}$ & $100.72^{+22.86}_{-52.86}$ & $3.76^{+0.92}_{-0.87}$ & 1.12 & $0.96^{+0.35}_{-0.24}$ & $1.47^{+0.48}_{-0.34}$ & $0.77^{+0.28}_{-0.20}$  & $0.13\tablenotemark{a}$ \\		
I 13 & 15.17 & $0.44^{+0.04}_{-0.02}$ & $129.10^{+51.62}_{-81.62}$ & $2.91^{+0.81}_{-0.73}$ & 1.12 & $1.46^{+0.54}_{-0.37}$ & $2.23^{+0.86}_{-0.51}$ & $0.99^{+0.41}_{-0.27}$  & $0.13\tablenotemark{a}$ \\		
I 14 & 15.86 & $0.46^{+0.03}_{-0.02}$ & $166.82^{+101.41}_{-111.41}$ & $2.96^{+1.11}_{-0.71}$ & 1.24 & $1.28^{+0.51}_{-0.32}$ & $2.14^{+0.77}_{-0.49}$ & $0.77^{+0.33}_{-0.22}$ & $0.34^{+0.32}_{-0.24}$ \\
I 15 & 16.79 & $0.60^{+0.06}_{-0.05}$ & $42.56^{+51.03}_{-18.03}$ & $2.31^{+0.61}_{-0.28}$ & 1.45 & $1.62^{+0.65}_{-0.42}$ & $3.14^{+1.21}_{-0.74}$ & $0.91^{+0.37}_{-0.24}$ & $0.24^{+0.24}_{-0.18}$ \\
I 16 & 17.96 & $0.82^{+0.13}_{-0.10}$ & $23.07^{+13.96}_{-8.25}$ & $1.85^{+0.49}_{-0.50}$ & 1.22 & $1.85^{+0.86}_{-0.54}$ & $4.34^{+2.15}_{-1.18}$ & $1.06^{+0.54}_{-0.31}$  & $0.13\tablenotemark{a}$ \\		
I 17 & 19.09 & $0.79^{+0.09}_{-0.06}$ & $20.32^{+18.91}_{-9.45}$ & $4.45^{+0.59}_{-0.57}$ & 1.24 & $0.47^{+0.16}_{-0.14}$ & $1.13^{+0.30}_{-0.28}$ & $0.30^{+0.10}_{-0.05}$  & $0.13\tablenotemark{a}$ \\		
I 18 & 20.56 & $0.56^{+0.04}_{-0.04}$ & $7.27^{+2.84}_{-2.90}$ & $6.83^{+0.90}_{-0.82}$ & 1.11 & $0.17^{+0.07}_{-0.05}$ & $0.22^{+0.06}_{-0.05}$ & $0.13^{+0.05}_{-0.04}$  & $0.13\tablenotemark{a}$ 
\enddata
\tablecomments{Abundances are with respect to solar \citep{ande89}. The Galactic column $N_{\rm H,Gal}$ is fixed at 0.45 $\times$ 10$^{21}$ cm$^{-2}$ and the SMC column $N_{\rm H,SMC}$ is fixed at the Shell value 0.8 $\times$ 10$^{21}$ cm$^{-2}$. Fe abundance was fixed at the best-fit Shell value. For comparisons the Russell \& Dopita (1992) values for SMC abundances are O = 0.126, Ne = 0.151, Mg = 0.251, Si = 0.302, Fe = 0.149}
\tablenotetext{a}{Si abundance was fixed at the best-fit Shell value}
\tablenotetext{b}{Two shock model parameters for transition region I3}
\end{deluxetable*}
\end{document}